\documentclass[aps,prd,notitlepage,nofootinbib,superscriptaddress,
groupedaddress,showpacs]{revtex4-1}
\usepackage{amsmath}
\usepackage{amssymb,amsfonts}
\usepackage{graphicx}
\usepackage[utf8]{inputenc}
\usepackage{color}
\usepackage{algorithm}
\usepackage{algorithmic}
\usepackage{listings}
\usepackage{textcomp}
\usepackage{url}
\usepackage{diagbox}
\usepackage{subfigure}
\usepackage{mathtools}
\usepackage[section]{placeins}

\newcommand{\ch}{\mathcal H}
\newcommand{\lt}{\left}
\newcommand{\rt}{\right}

\newcommand{\be}{\begin{equation}}
\newcommand{\ee}{\end{equation}}
\newcommand{\dd}{\mathrm{d}}
\newcommand{\bse}{\begin{subequations}}
\newcommand{\ese}{\end{subequations}}

\newcommand{\bracket}[2]{\langle{#1}\vert{#2}\rangle}
\newcommand{\Z}{\mathbb{Z}}
\newcommand{\ii}{\mathrm{i}}
\newcommand{\e}{\mathrm{e}}

\newcommand{\bpm}{\begin{pmatrix}}
\newcommand{\epm}{\end{pmatrix}}
\newcommand{\bmm}{\begin{matrix}}
\newcommand{\emm}{\end{matrix}}

\newcommand{\x}{\times}

\allowdisplaybreaks

\makeatletter
\newcommand*{\Relbarfill@}{\arrowfill@\Relbar\Relbar\Relbar}
\newcommand*{\xeq}[2][]{\ext@arrow 0055\Relbarfill@{#1}{#2}}
\makeatother

\newcommand{\bra}[1]{\langle {#1}|}
\newcommand{\ket}[1]{| #1 \rangle}

\newcommand{\half}{\frac{1}{2}}

\newcommand{\R}{\mathbb{R}}

\usepackage[normalem]{ulem}
\renewcommand\sout{\bgroup\markoverwith{\textcolor{red}{\rule[0.5ex]{4pt}{0.8pt}}}\ULon}

\def\mb{\left(\begin{matrix}}
\def\me{\end{matrix}\right)}

\newcommand{\Psix}[3][1]{
\begin{tikzpicture}[scale=0.8]
\node[name=s, regular polygon, regular polygon sides=6, minimum size=1cm, outer sep=0pt ,draw] at (0,0) {}; 
\foreach \anchor/\x/\y /\xx/\yy /\b in
{corner 1/0.17/0.17*1.732/-0.11/0.18/1, corner 2/-0.17/0.17*1.732/0.07/0.18/2, corner 3/-0.34/0/-0.15/-0.18/3, corner 4/-0.17/-0.17*1.732/-0.22/-0.05/4, corner 5/0.17/-0.17*1.732/0.2/-0.05/5, corner 6/0.34/0/0.15/-0.18/6}
{
 \draw[shift=(s.\anchor)] (0,0) -- (\x,\y) node at(\xx,\yy) {$#2_{\text{\scalebox{0.7}{$\b$}}}$};
 \ifnum #1=1
 \draw[shift=(s.\anchor),<-,>=stealth', line width=0.01pt] (s.\anchor) -- (\x,\y);
 \fi
 }
\foreach \anchor/\xx/\yy /\a in
{side 1/0/-0.18/1, side 2/-0.18/0.05/2, side 3/0.15/0.05/3, side 4/0/-0.18/4, side 5/-0.18/0.05/5, side 6/0.15/0.05/6}
 \draw[shift=(s.\anchor)]  node at(\xx,\yy) {$#3_{\text{\scalebox{0.7}{$\a$}}}$};
\ifnum #1=1{
  \foreach \anchorr/\anchorf in
   {corner 1/corner 2, corner 2/corner 3, corner 3/corner 4, corner 4/corner 5, corner 5/corner 6, corner 6/corner 1}
   \draw[shift=(s.\anchorr), ->, >=stealth', line width=0.01pt]  (s.\anchorr) -- (s.\anchorf);}
 \else {
  \foreach \anchorb/\anchorw in
   {corner 1/corner 2, corner 3/corner 4, corner 5/corner 6} {
   \node[fill=black, circle, minimum size=0, inner sep=0, outer sep=0, draw] at(s.\anchorb) {};
   \node[fill=white, circle, minimum size=0, inner sep=0, outer sep=0, draw] at(s.\anchorw) {};}
}
\fi
\end{tikzpicture}
}

\usepackage{varioref}

\begin{document}

\title{Spinfoam on Lefschetz Thimble: Markov Chain Monte-Carlo Computation of Lorentzian Spinfoam Propagator}
\author{\ Muxin Han}  
\affiliation{Department of Physics, Florida Atlantic University, 777 Glades Road, Boca Raton, FL 33431-0991, USA}
\affiliation{Institut f\"ur Quantengravitation, Universit\"at Erlangen-N\"urnberg, Staudtstr. 7/B2, 91058 Erlangen, Germany}

\author{\ Zichang Huang}
\email{hzc881126@hotmail.com}  
\affiliation{State Key Laboratory of Surface Physics, Fudan University, Shanghai 200433, China}
\affiliation{Department of Physics, Center for Field Theory and Particle Physics, and Institute for Nanoelectronic devices and Quantum computing, Fudan University, Shanghai 200433, China}
\author{\ Hongguang Liu}
\affiliation{Institut f\"ur Quantengravitation, Universit\"at Erlangen-N\"urnberg, Staudtstr. 7/B2, 91058 Erlangen, Germany}

\author{\ Dongxue Qu}
\affiliation{Department of Physics, Florida Atlantic University, 777 Glades Road, Boca Raton, FL 33431-0991, USA}

\author{\ Yidun Wan}
\affiliation{State Key Laboratory of Surface Physics, Fudan University, Shanghai 200433, China}
\affiliation{Department of Physics, Center for Field Theory and Particle Physics, and Institute for Nanoelectronic devices and Quantum computing, Fudan University, Shanghai 200433, China}
\affiliation{Department of Physics and Institute for Quantum Science and Engineering, Southern University of Science and Technology, Shenzhen 518055, China}

\begin{abstract}

We compute numerically the Lorentzian Engle-Pereira-Rovelli-Livine (EPRL) spinfoam propagator on a 4-simplex, by adapting the methods of Lefschetz thimble and Markov Chain Monte-Carlo to oscillatory spinfoam integrals. Our method can compute any spinfoam observables at relatively large spins. We obtain the numerical results of the propagators at different spins and demonstrate their consistency with the expected spinfoam semi-classical behavior in the large spin limit. Our results exhibit significant quantum corrections at smaller spins. Our method is reliable and thus can be employed to discover the semi-classical and quantum behaviors of the spinfoam model.

\end{abstract}

\date{\today}
\maketitle


\section{Introduction}
Computing and studying properties of the $n$-point correlation functions are the core problem in quantum field theory. As emphasized by Arthur Wightman, a complete quantum field theory can be uniquely reconstructed from its $n$-point correlation functions. The $n$-point functions in Loop Quantum Gravity (LQG) \cite{book,carlobook,review,review1}, which is a promising candidate of the quantum gravity theory, have been introduced in \cite{propagator,3pt}. In \cite{propagator1,propagator2,propagator3,Han:2017isy}, analysis has been carried out on the 2-point correlation function for the Penrose metric operators in the spinfoam model---the covariant formulation of LQG \cite{EPRL,Freidel:2007py,Reisenberger:2000fy,Perez:2003vx,carlobook}. This 2-point functions are usually called the spinfoam propagator. The analysis confirms that the semiclassical behavior of the spinfoam propagators, determined by the large-spin asymptotics, matches the one obtained from the Regge calculus, and indicates that LQG recovers General Relativity in 4 dimensions in an appropriate limit. Asymptotic analysis of the spinfoam amplitudes have drawn similar conclusion\cite{Barrett:2009mw,BARRETT_2010,CFsemiclassical,LowE1,HZ,HZ1,Kaminski:2017eew,Liu_2019,Han:2018fmu,Dona:2020yao,HHKR}. Despite the existence of the novel and crucial results brought in by these discussions of spinfoam propagators, the computational complexity has been obstructed further explorations in the spinfoam model. Nevertheless, numerical approaches to spinfoams open new windows to circumvent this obstruction. The works \cite{Dona:2019dkf,dona2020numerical,Don__2018} have attempted numerical computing of the spinfoam amplitudes and obtained enlightening results. There are also numerical results on the spinfoam renormalization \cite{Bahr:2016hwc,Bahr:2018gwf}. In early attempts, the numerical computations of th spinfoam correlation function for 3d gravity \cite{Speziale:2005ma} and 4d Barrett–Crane model \cite{CHRISTENSEN2009403,Christensen_2010} are done. Yet, the numerical computation of the spinfoam correlation function for Engle-Pereira-Rovelli-Livine (EPRL) model has not been well developed.

In this paper, we propose and employ a numerical method that combines Lefschetz thimble and Markov Chain Monte-Carlo to compute the spinfoam propagator on a 4-simplex in the EPRL spinfoam model. In contrast to the algorithm in \cite{Dona:2019dkf,dona2020numerical,Don__2018} which mostly applicable to spinfoam amplitudes with small spins, our algorithm is featured by computing spinfoams with relatively large spins. We obtain the numerical results of the propagators at different spins and compare with the spinfoam large spin asymptotics. Our numerical results are shown to be consistent with the spinfoam asymptotics in the large spin limit while providing significant corrections at smaller spins. 

In the path integral formalism, the expectation value of an arbitrary observable $\mathcal{O}[\varphi]$ can be expressed as 
\begin{equation}\label{eq:1st}
\langle \hat{\mathcal{O}} \rangle =\frac{\int D\varphi\, \mathcal{O}[\varphi] \,\e^{-S[\varphi]}}{\int D\varphi\, \e^{-S[\varphi]}},
\end{equation}
where $S$ is the action. The spinfoam propagator can be expressed in terms of similar integral expressions \cite{propagator3}
\be
G^{a b c d}_{mn}=\left\langle E^a_n\cdot E^b_n E^c_m\cdot E^d_m \right\rangle-\left\langle E^a_n\cdot E^b_n\right\rangle\left\langle E^c_m\cdot E^d_m\right\rangle,
\ee
where each $E^a_n$ is the flux operator at the face shared by tetrahedra $n$ and $a$. The action $S$ for the propagator is a complex valued function depending on 54 real variables (for one 4-simplex). Both the denominator and the nominator in the right hand side of Eq. \eqref{eq:1st} are integrals of high-dimensional oscillatory functions. Known as the sign problem \cite{PhysRevD.66.106008,deforcrand2010simulating,Philipsen:2012qt}, these oscillatory integrals cannot be evaluated in the conventional Monte Carlo integral method, thus have been difficult to be studied numerically. 

In order to compute $\left\langle O \right\rangle$, recent progresses (see e.g.  \cite{Witten:2010cx,Witten:2010zr,Alexandru:2020wrj,Cristoforetti:2013qaa,Alexandru:2017czx,Alexandru:2015xva}) suggest to apply the \textit{Picard-Lefschetz} theory to cure the sign problem, and have applied the theory to numerical computations in Lattice Field Theories. In our work, we apply the theory to the numerical computation of the spinfoam propagator and improve the numerical method for handling higher dimensional oscillatory integrals. The idea of the method is to deform the integration contour to a class of specific integral cycles, called \textit{Lefschetz thimbles} $\mathcal{J}_\sigma$, each of which is defined as the union of all steepest descent paths ending at a given critical point of the analytic continued action. The deformation does not change the value of the integral, while it has the advantage that the imaginary part of $S$ is constant on each Lefschetz thimble $\mathcal{J}_\sigma$, so $\e^{-S}$ becomes non-oscillatory on the thimble. The integral on each $\mathcal{J}_\sigma$ can be studied numerically with the Monte-Carlo method.

Technically, the numerical integration on the Lefschetz thimble $\mathcal{J}_\sigma$ can be achieved by the Markov Chain Monte Carlo (MCMC) method \cite{brooks_meng_jones_galman_2011} \footnote{See \cite{Barrau:2011md,Steinhaus:2018aav} for some existing works on applying Monte Carlo method to LQG/Spinfoam.}, which treats the $\left\langle O \right\rangle$ on the thimble as the mean value of $O e^{i\arg(\det J)}$ among samples on $\mathcal{J}_\sigma$ given by the Boltzmann distribution $\e^{-{\rm Re}(S)+\log|\det J|}$ where $\det J$ is the measure factor on $\mathcal{J}_\sigma$. Specifically, in order to do the numerical integration on a high-dimensional thimble, we use a multi-chain MCMC method called the Differential Evolution Adaptive Metropolis (DREAM) algorithm \cite{DREAM} to compute the $\left\langle O \right\rangle$ on the thimble.


The stationary phase analysis shows that in the large spin regime, $G^{a b c d}_{mn}$ receives the dominant contribution from a single critical point corresponding to the Lorentizian geometrical 4-simplex. This infers that for large spins, the Lefschetz thimble attached to this geometrical critical point dominates the integrals for computing the propagator, while it turns out that the Lefschetz thimbles of other critical point only give exponentially small contributions.

In the case where the contribution to $\left\langle O \right\rangle$ of certain thimble is much larger than the contributions of the other thimbles, one can compute the integral on the dominant thimble and use it as a proper approximation of $\left\langle O \right\rangle$. The integral on the Lefschetz thimble can generate the same perturbative expansion as the stationary phase analysis \cite{Cristoforetti:2012su}. Thus this approximation keeps all the \emph{perturbative} quantum corrections by integrating along the dominant thimble, while neglecting \emph{non-perturbative} corrections from other thimbles whose contributions are exponentially suppressed. 

In our computation, the operators expectation values $\left\langle E^a_n\cdot E^b_n E^c_m\cdot E^d_m \right\rangle$, $\left\langle E^a_n\cdot E^b_n\right\rangle$, $\left\langle E^c_m\cdot E^d_m\right\rangle$ are computed numerically on the dominant thimble attached to the geometrical critical point. This computation capture perturbative quantum corrections to all orders in the spinfoam model, while neglecting non-perturbative corrections that are exponentially small for large spins. Given that the computational complexity is scaled exponentially large when computing perturbative expansion to high orders, our method with Lefschetz thimble is efficient and powerful. 


In practice, we compute the spinfoam propagator with the coherent boundary state peaked at the boundary data of a Lorentizian geometrical 4-simplex. We scale the boundary spins by $\lambda$ and numerically compute the propagator with $\lambda\in [10^2,5\times10^7]$. In the regime $\lambda>10^4$, our numerical result strongly tends to comply with the expected semiclassical limit. At smaller spins, the numerical results provide significant quantum corrections. This consistency supports the reliability of our algorithm and increases our confidence to use the algorithm as a tool to discover the semi-classical and quantum behaviors of the spinfoam model.

We emphasize that our method applies to a much boarder range of research areas involving the complex-valued actions and oscillatory integrals, e.g., finite density lattice QCD, Lee-Yang zeros, non-Hermitian systems, gauge theory with topological $\theta$-term, etc. Similar methods have been applied in these areas \cite{Alexandru:2020wrj,Kanazawa2015,Tanizaki_2016,fukuma2020worldvolume,PhysRevD.91.101701,PhysRevD.101.014508,Ulybyshev:2019hfm}. We look forward to finding novel interesting results by applying our method to these areas.   

This paper is organized as follows: Section II discusses the Lorentzian spinfoam propagator in the EPRL model, the boundary state, and our parameterizations of the action the observables. Section III explains our algorithm, including reviewing the Picard-Lefschetz theory, MCMC, and DREAM. Section IV discusses the application of Lefschetz thimble to the spinfoam model and a few conceptual aspects. Section V discusses several optimizations to improve the efficiency of the computation. Section VI discusses the stationary phase approximation and the large spin limit of the spinfoam propagator. Section VII presents our numerical results from the Monte-Carlo on Lefschetz-thimble and compares with the large spin limit. Section VIII presents the benchmarks the platforms that we use for testing our code. In Section IX, we conclude and mention a few future perspectives.

We tested our codes in \textit{Mathematica}\texttrademark 12. The codes are posted on \cite{hzcgit}.


\section{Spinfoam Propagator}\label{sec:SFP}

In this section, we review the EPRL spinfoam model, including the definition of the vertex amplitude and the definition of the spinfoam propagator, and we setup the boundary data and critical point used in the numerical computation. In the integral representation of the spinfoam vertex amplitude, the integrand is a function depending on $54$ real variables. Furthermore, we present four integrals to be computed by our numerical algorithm for computing the spinfoam propagator. 

\subsection{Boundary State}\label{lab:bs}
The boundary state is crucial in calculating the spinfoam amplitude. Following the basic setup in \cite{propagator3}, our boundary state defined below is a \textit{Lorentzian semi-classical state} peaked at both intrinsic and extrinsic geometry. We consider a 4-dimensional Lorentzian spacetime region $\mathcal{R}$ homeomorphic to a 4-ball whose 3-dimensional boundary $\partial {\mathcal{R}}$ is homeomorphic to a 3-sphere. Taking the coarsest triangulation, $\mathcal{R}$ is a 4-simplex, and $\partial {\mathcal{R}}$ is triangulated by five tetrahedra. Specifically, in this paper, we use the boundary geometry introduced in \cite{Dona:2019dkf,Han:2020fil}. Namely, we set the five vertices of the 4-simplex as 
\[
  \begin{split}
    P_1=(0, 0, 0, 0),\ 
    P_2=(0, 0, 0, -2 &\sqrt{5}/3^{1/4}),\ 
    P_3=(0, 0, -3^{1/4} \sqrt{5},-3^{1/4} \sqrt{5}),\\
    P_4=(0, -2 \sqrt{10}/3^{3/4}, -\sqrt{5}/3^{3/4}, -\sqrt{5}/3^{1/4}),\ 
    &P_5=(-3^{-1/4} 10^{-1/2}, -\sqrt{5/2}/3^{3/4}, -\sqrt{5}/3^
  {3/4}, -\sqrt{5}/3^{1/4}). 
  \end{split}
\]
Then, the 4-dimensional normal vectors of the tetrahedra\footnote{Here, the we use $i\ (=1,\dots,5)$ to label the tetrahedron that does not contain the vertex $P_i$.} are given by:
\begin{equation}\label{eq:4Normals}
\begin{split}
N_1=&\left(-1,0,0,0\right),\ 
N_2=\left(\frac{5}{\sqrt{22}},\sqrt{\frac{3}{22}},0,0\right),\ 
N_3=\left(\frac{5}{\sqrt{22}},-\frac{1}{\sqrt{66}},\frac{2}{\sqrt{33}},0\right),\\ 
N_4&=\left(\frac{5}{\sqrt{22}},-\frac{1}{\sqrt{66}},-\frac{1}{\sqrt{33}},\frac{1}{\sqrt{11}}\right),\ 
N_5=\left(\frac{5}{\sqrt{22}},-\frac{1}{\sqrt{66}},-\frac{1}{\sqrt{33}},-\frac{1}{\sqrt{11}}\right).
\end{split} 
\end{equation}

The areas of the 10 faces are uniformly proportional to the data shown in Table \ref{tab:facearea}, and the 3-dimensional normal of each face\footnote{We use the pair $(ab)$ to indicate the face of the tetrahedron $a$ pointing to tetrahedron $b$.} is shown in Table \ref{tab:3dNormal}. 
\begin{table}[h]
	\centering\caption{Each cell shows the area of the face shared by line number tetrahedra and column number tetrahedra.}\label{tab:facearea}
	\small
	\setlength{\tabcolsep}{0.8mm}
	\begin{tabular}{|c|c|c|c|c|}
    \hline
		\diagbox{\small{a}}{area $j_0{}_{ab}$}{\small{b}}&2&3&4&5\\
		\hline
		1&5&5&5&5\\
		\hline
		2&\diagbox{}{}&2&2&2\\
		\hline
		3&\diagbox{}{}&\diagbox{}{}&2&2\\
		\hline
		4&\diagbox{}{}&\diagbox{}{}&\diagbox{}{}&2\\
		\hline
	\end{tabular}
\end{table} 

\begin{table}[h]
	\centering\caption{Each cell shows the 3-dimensional normal vector of the face shared by line number tetrahedra and column number tetrahedra.}\label{tab:3dNormal}
	\small
	\setlength{\tabcolsep}{0.8mm}
  \begin{tabular}{|c|c|c|c|c|c|}
    \hline
		\diagbox{\small{a}}{normal $\vec{n}_{ab}$ }{\small{b}}&1&2&3&4&5\\
		\hline
		1&\diagbox{}{}&(1,0,0)&(-0.33,0.94,0)&(-0.33,-0.47,0.82)&(-0.33,-0.47,-0.82)\\
		\hline
		2&(-1,0,0)&\diagbox{}{}&(0.83,0.55,0)&(0.83,-0.28,0.48)&(0.83,-0.28,-0.48)\\
		\hline
		3&(0.33,-0.94,0)&(0.24,0.97,0)&\diagbox{}{}&(-0.54,0.69,0.48)&(-0.54,0.69,-0.48)\\
		\hline
		4&(0.33,0.47,-0.82)&(0.24,-0.48,0.84)&(-0.54,0.068,0.84)&\diagbox{}{}&(-0.54,-0.76,0.36)\\
		\hline
		5&(0.33,0.47,0.82)&(0.24,-0.48,-0.84)&(-0.54,0.068,-0.84)&(-0.54,-0.76,-0.36)&\diagbox{}{}\\
		\hline
	\end{tabular}
\end{table} 

The intrinsic and extrinsic geometry of the boundary tetrahedra are given by the face areas, 3d normals, and 4d normals. The corresponding semi-classical boundary quantum state is a superposition of coherent spin-network states
\be
\ket{\Psi_0}=\sum_{\lambda j_{ab}}\psi_{\lambda j_0,\zeta_0}|\ket{\lambda j_{ab},\vec{n}_{ab}}.
\ee
We have denoted the spins by $\lambda j_{ab}$ and $\lambda j_0{}_{ab}$ where $j_0{}_{ab}$ are recorded in Table \ref{tab:facearea} and $\lambda$ is a scaling parameter. The large spin limit corresponds to large $\lambda$. The five \textit{Livine-Speziale coherent intertwiner} $|\ket{\lambda j_{ab},\vec{n}_{ab}}$ are compatible with the $\vec{n}_{ab}$ in Table \ref{tab:3dNormal}. The wave packet $\psi_{j_0,\zeta_0}$ reads
\[
  \psi_{\lambda j_0,\zeta_0}=\exp\left( -i \sum_{ab} \zeta_0^{ab}(\lambda j_{ab}-{\lambda j_{0}}_{ab})\right) 
  \exp\left(-\sum_{ab,cd}\alpha^{(ab)(cd)} \frac{\lambda j_{ab}-{\lambda j_{0}}_{ab}}{\sqrt{{\lambda j_{0}}_{ab}}}\frac{\lambda j_{cd}-{\lambda j_{0}}_{cd}}{\sqrt{{\lambda j_{0}}_{cd}}}\right),
\]
where $\zeta_0^{ab}$, whose values are given in TABLE \ref{tab:zeta}, is related to the dihedral angles between the 4-normals \eqref{eq:4Normals} (see Section \ref{Large Spin Approximation} for the way to determine $\zeta_0^{ab}$). The spin variables $\lambda {j_{0}}_{ab}$ correspond to the areas listed in TABLE \ref{tab:facearea}. The matrix $\alpha^{(ab)(cd)}$ has positive definite real part, and  
\[
  \alpha^{(ab)(cd)}=\alpha_1P_0^{(ab)(cd)}+\alpha_2P_1^{(ab)(cd)}+\alpha_3P_2^{(ab)(cd)},
\]
where $\alpha_1,\alpha_2,\alpha_3$ are free parameters, and $P_k^{(ab)(cd)}\ (k=0\cdots2)$ are defined as
\begin{itemize}
  \item $P_0^{(ab)(cd)}=1$ if $(ab)=(cd)$ and zero otherwise;
  \item $P_1^{(ab)(cd)}=1$ if $a=c,\ b\neq d$ and zero otherwise;
  \item $P_2^{(ab)(cd)}=1$ if $(ab)\neq(cd)$ and zero otherwise.
\end{itemize}
The spinfoam amplitude with coherent spin-networks as the boundary state depends on the free parameters $\alpha$. In our numerical computation, we set $\alpha_1=0.2,\alpha_2=0.3,\alpha_3=0.4$ for definiteness. Any other choice of $\alpha$ does not affect the application of our algorithm. 



\begin{table}[h]
	\centering\caption{The table of $\zeta^{ab}_0$}\label{tab:zeta}
	\small
	\setlength{\tabcolsep}{0.8mm}
  \begin{tabular}{|c|c|c|c|c|c|}
    \hline
		\diagbox{\small{a}}{ $\zeta^{ab}_0$ }{\small{b}}&2&3&4&5\\
		\hline
		1&-3.14+0.36$\gamma$&0.68+0.36$\gamma$&5.05+0.36$\gamma$&5.05+0.36$\gamma$\\
		\hline
		2&\diagbox{}{}&5.05-0.59$\gamma$&-5.93-0.59$\gamma$&-3.20-0.59$\gamma$\\
		\hline
		3&\diagbox{}{}&\diagbox{}{}&-2.81-0.59$\gamma$&-5.54-0.59$\gamma$\\
		\hline
		4&\diagbox{}{}&\diagbox{}{}&\diagbox{}{}&-4.37-0.59$\gamma$\\
		\hline
	\end{tabular}
\end{table} 



\subsection{Spinfoam Action}
In the boundary formalism \cite{oeckl2003general,Oeckl_2008,carlobook}, the spinfoam amplitude for the boundary state $\ket{\Psi_0}$ can be written as 
\be
\bracket{W}{\Psi_0}=\sum_{j_{ab}}\psi_{\lambda j_0,\zeta_0}\bra{W}\ket{\lambda j_{ab},\vec{n}_{ab}},
\ee
where $\bra{W}$ is a $\mathbb{C}$-valued linear functional providing the sum over the bulk geometries with the weight that defines our model for quantum gravity. The Lorentzian EPRL vertex amplitude can be expressed as
\be
\bracket{W}{\Psi_0}=\sum_{j_{ab}}\psi_{j_0,\zeta_0}\int_{S L(2, \mathbb{C})^{5}} \prod_{a} \mathrm{d} g_{a} \prod_{a>b} P_{a b}(g),
\ee
with 
\be\label{eq:brac}
P_{a b}(g)=\left\langle \lambda j_{a b},-\vec{n}_{a b}\left|Y^{\dagger} g_{a}^{-1} g_{b} Y\right| \lambda j_{a b}, \vec{n}_{b a}\right\rangle.
\ee
In \eqref{eq:brac}, $Y$ maps the spin-$j$ $SU(2)$ irreducible representation $\mathcal{H}_j$ to the lowest level in $SL(2,\mathbb{C})$ $(j,\gamma j)$-irreducible representation $\mathcal{H}_{(j,\gamma j)}=\oplus_{k=j}^\infty\ch_k$ (See \cite{Ding:2010fw,Dona:2020xzv} for alternative choices of the $Y$-map.)\footnote{Our numerical algorithm is applicable to these alternative choices of the $Y$-map.}. Using the fact that the elements in $\mathcal{H}_{(j,\gamma j)}$ can be expressed as homogeneous functions on $\mathbb{CP}_1$, the inner product \eqref{eq:brac} is equivalent to an integral \cite{Barrett:2009mw,HZ}
\be
P_{ab}=\frac{d_{\lambda j_{a b}}}{\pi} \int_{\mathbb{C P}^{1}} \mathrm{d} \tilde{\mathbf{z}}_{a b}\left\langle Z_{b a}, Z_{b a}\right\rangle^{-(1-\ii \gamma)\lambda j_{ab}}\left\langle Z_{a b}, Z_{a b}\right\rangle^{-(1+\ii \gamma )\lambda j_{a b}}\left\langle J \xi_{a b}, Z_{a b}\right\rangle^{2 \lambda j_{a b}}\left\langle Z_{b a}, \xi_{b a}\right\rangle^{2\lambda j_{a b}},
\ee 
in which $d_j=2j+1$, $Z_{a b} = g_{a}^{\dagger} z_{ab}$, and $Z_{b a} = g_{h}^{\dagger} z_{ab}$. The integral measure $\mathrm{d} \tilde{\mathbf{z}}_{a b} = -\left(\left\langle Z_{a b}, Z_{a b}\right\rangle\left\langle Z_{b a}, Z_{b a}\right\rangle\right)^{-1} \mathrm{d} \mathbf{z}_{ab}$ (with $\mathrm{d} \mathbf{z}=\frac{i}{2}\left(z_{0} d z_{1}-z_{1} d z_{0}\right) \wedge\left(\bar{z}_{0} d \bar{z}_{1}-\bar{z}_{1} d \bar{z}_{0}\right)$) is homogeneous on $\mathbb{C P}^{1}$. The bracket $\langle\ ,\ \rangle$ is the Hermitian inner product on $\mathbb{C}^2$. The spinors $\xi_{ba}$ and $J\xi_{ab}$ \footnote{For a spinor $Z=(z_1,z_2)$, $JZ=(\bar{z}_2,-\bar{z}_1)$. } are related to the 3-normal $\vec{n}_{ba}$ and $-\vec{n}_{ba}$ respectively by $\vec{n}_{ba}=\langle \xi_{ba}|\vec{\sigma}|\xi_{ba}\rangle$ and $-\vec{n}_{ab}=\langle J\xi_{ab}|\vec{\sigma}|J\xi_{ab}\rangle$. TABLE \ref{tab:xi} lists the spinors that compatible with our boundary geometry.
\begin{table}[h]
  \centering\caption{Each cell indicates a spinor $\xi_{ab}$ corresponding to a 3-normal of a tetrahedron.}\label{tab:xi}
\footnotesize
\setlength{\tabcolsep}{0.8mm}
\begin{tabular}{|c|c|c|c|c|c|}
  \hline
  \diagbox{\small{a}}{$\ket{\xi_{ab}}$ }{\small{b}}&1&2&3&4&5\\
  \hline
  1&\diagbox{}{}&(0.71,0.71)&(0.71,-0.24+0.67 i)&(0.95,-0.17-0.25 i)&(0.30,-0.55-0.78 i)\\
  \hline
  2&(0.71,-0.71)&\diagbox{}{}&(0.71,0.59+0.39 i)&(0.86, 0.48 - 0.16 i)&(0.51, 0.82 - 0.27 i)\\
  \hline
  3&(0.71, 0.24 - 0.67 i)&(0.71, 0.17 + 0.69 i)&\diagbox{}{}&(0.86, -0.31 + 0.40 i)&(0.51, -0.53 + 0.68 i)\\
  \hline
  4&(0.30, 0.55 + 0.78 i)&(0.96, 0.13 - 0.25 i)&(0.96, -0.28 + 0.035 i)&\diagbox{}{}&(0.83, -0.33 - 0.46 i)\\
  \hline
  5&(0.95, 0.17 + 0.25 i)&(0.28, 0.43 - 0.86 i)&(0.28, -0.95+ 0.12 i)&(0.57, -0.48-0.67 i)&\diagbox{}{}\\
  \hline
\end{tabular}
\end{table} 

Using the integral expression of $P_{ab}$, we can write the amplitude as 
\be\label{eq:amp1}
\bracket{W}{\Psi_0}=\sum_{\lambda j_{ab}}\psi_{\lambda j_0,\zeta_0}\bra{W}\ket{\lambda j_{ab},\vec{n}_{ab}}=\sum_{\lambda j_{ab}}\psi_{\lambda j_0,\zeta_0}\int_{S L(2, \mathbb{C})^{5}} \prod_{a} \mathrm{d} g_{a}\int\left(\prod_{a>b} \frac{d_{\lambda j_{a b}}}{\pi} \mathrm{d} \tilde{\mathbf{z}}_{a b}\right) \e^{\lambda S},
\ee
with the spinfoam action $S$ (without the scaling parameter $\lambda$) given by 
\be\label{eq:action1}
  S(j, g, \mathbf{z})=\sum_{a>b} \left[2 j_{a b} \log(\left\langle J \xi_{a b}, Z_{a b}\right\rangle\left\langle Z_{b a}, \xi_{b a}\right\rangle)
  -(1+\ii\gamma)j_{ab}\log\left\langle Z_{a b}, Z_{a b}\right\rangle-(1-\ii\gamma) j_{ab} \log\left\langle Z_{b a}, Z_{b a}\right\rangle\right].
\ee 
After gauge fixing $g_1=1$, the action is a function of four $SL(2, \mathbb{C})$ elements $g_a,\ (a=2\cdots5 )$, ten spinors $z_{ab}$, and ten area variables $j_{ab}$. Explicitly, we parametrize each $SL(2, \mathbb{C})$ element by six real parameters as
\be\label{eq:par1}
g=\left(
\begin{matrix}
  1+\frac{x_1+\ii y_1}{\sqrt{2}} && \frac{x_2+\ii y_2}{\sqrt{2}}\\
  \frac{x_3+\ii y_3}{\sqrt{2}} && \frac{1+\frac{x_2+\ii y_2}{\sqrt{2}}\frac{x_3+\ii y_3}{\sqrt{2}}}{1+\frac{x_1+\ii y_1}{\sqrt{2}}}\\
\end{matrix}
\right),
\ee
and we parametrize each spinor by two real parameters in a way that
\be\label{eq:par2}
z=(1,x+\ii y).
\ee
Each group variable is parametrized by six real parameters, and each spinor variable is parametrized by two real parameters. As such, the action $S$ is a function depending on $54$ real parameters. 

In this parametrization, the measure of the spinor $dz$ becomes
\[
\mathrm{d}\mathbf{z}=dx dy, 
\]
and the Haar measure $dg$ of $SL(2, \mathbb{C})$ is expressed as (see Appendix A in \cite{Han:2020fil}) 
\[
dg=\frac{1}{128\pi^4}\frac{dx_1dx_2dx_3dy_1dy_2dy_3}{|1+\frac{x+\ii y}{\sqrt{2}}|^2}.
\]
The amplitude \eqref{eq:amp1} is indeed a superposition of multiple $44$-dimensional integrals
\be\label{eq:core0}
\bracket{W}{\Psi_0}=\sum_{j_{ab}}\psi_{\lambda j_0,\zeta_0}\int \mathrm{d} \phi\, U(j,\phi)\, \e^{\lambda S(j,\phi)},
\ee
where $\phi$ stands for $44$ real variables parametrizing $g$ and $z$, and 
\[
U(j, \phi)=  \frac{1}{(128\pi^4)^4} \left(\prod_{a>b} -\frac{d_{\lambda j_{a b}}}{\pi} 
\left(\left\langle Z_{a b}, Z_{a b}\right\rangle\left\langle Z_{b a}, Z_{b a}\right\rangle\right)^{-1} \right)
\prod_{a} \frac{1}{|(g_a)_{1,1}|^2}.
\]

\subsection{Spinfoam Propagator}\label{subsec:SFP}
Following the boundary formalism, the expectation value of an observable $\hat{\mathcal{O}}$ is defined by
\[
\langle \hat{\mathcal{O}} \rangle=\frac{\bra{W}\hat{\mathcal{O}}\ket{\Psi_0}}{\bracket{W}{\Psi_0}}.  
\]
The spinfoam propagator $G^{a b c d}_{mn}$ is constructed as \cite{propagator3,propagator2,propagator1,propagator}
\be
\begin{split}\label{eq:prop1}
G^{a b c d}_{mn}=&\frac{\bra{W}E^a_n\cdot E^b_n E^c_m \cdot E^d_m |\Psi_0\rangle}{\bracket{W}{\Psi_0}}
-\frac{\bra{W}E^a_n\cdot E^b_n |\Psi_0\rangle}{\bracket{W}{\Psi_0}}\frac{\bra{W} E^c_m \cdot E^d_m |\Psi_0\rangle}{\bracket{W}{\Psi_0}}\\
=&\frac{\sum_{ j_{ab}}\psi_{\lambda j_0,\zeta_0}\bra{W}E^a_n\cdot E^b_n E^c_m \cdot E^d_m |\ket{\lambda j_{ab},\vec{n}_{ab}}}{\sum_{ j_{ab}}\psi_{\lambda j_0,\zeta_0}\bra{W}\ket{\lambda j_{ab},\vec{n}_{ab}}}\\
&-\frac{\sum_{ j_{ab}}\psi_{\lambda j_0,\zeta_0}\bra{W}E^a_n\cdot E^b_n |\ket{\lambda j_{ab},\vec{n}_{ab}}}{\sum_{ j_{ab}}\psi_{\lambda j_0,\zeta_0}\bra{W}\ket{\lambda j_{ab},\vec{n}_{ab}}}\frac{\sum_{ j_{ab}}\psi_{\lambda j_0,\zeta_0}\bra{W} E^c_m \cdot E^d_m |\ket{\lambda j_{ab},\vec{n}_{ab}}}{\sum_{ j_{ab}}\psi_{\lambda j_0,\zeta_0}\bra{W}\ket{\lambda j_{ab},\vec{n}_{ab}}}.
\end{split}
\ee
where $E^a_n\cdot E^b_n$ is the spatial metric operator at the $n$-th tetrahedron. By definition, each flux operator ${E^a_b}^i$ can only act on the corresponding face state $\ket{\lambda j_{ab}, \vec{n}_{ba}}$. Thus we can explicitly express $\bra{W}E^a_n\cdot E^b_n E^c_m \cdot E^d_m |\ket{\lambda j_{ab},\vec{n}_{ab}}$, $\bra{W}E^a_n\cdot E^b_n |\ket{\lambda j_{ab},\vec{n}_{ab}}$, and $\bra{W} E^c_m \cdot E^d_m |\ket{\lambda j_{ab},\vec{n}_{ab}}$ in integral forms by inserting ${E^a_b}^i$ into $P_{ab}$. Since $P_{ab}$ appears in the amplitude under the condition $a>b$\footnote{We assume $a$ is always greater than $b$ here.}, the operator $E^a_b$ and the operator $E_a^b$ are inserted in $P_{ab}$ in two different ways. When inserting ${E^a_b}^i$ into the right hand side of Eq. \eqref{eq:brac}, the $P_{ab}$ becomes 
\be
\begin{split}
{Q^a_b}^i \equiv&\left\langle \lambda j_{a b},-\vec{n}_{a b}\left|Y^{\dagger} g_{a}^{-1} g_{b} Y\left(E_{b}^{a}\right)^{i}\right| \lambda j_{a b}, \vec{n}_{b a}\right\rangle\\
=&\frac{d_{\lambda j_{a b}}}{\pi} \int_{\mathbb{C} \mathbb{P}^{1}} \mathrm{d} \tilde{\mathbf{z}}_{a b}\left\langle Z_{b a}, Z_{b a}\right\rangle^{-(1-\ii \gamma)\lambda j_{ab}}\left\langle Z_{a b}, Z_{a b}\right\rangle^{-(1+\ii \gamma )\lambda j_{a b}}\times\\
&\left\langle J \xi_{a b}, Z_{a b}\right\rangle^{2 \lambda j_{a b}}\left\langle Z_{b a}, \xi_{b a}\right\rangle^{2\lambda j_{a b}}\lambda j_{a b} 
\gamma \frac{\left\langle  \sigma_{i}Z_{ba}, \xi_{b a}\right\rangle}{\left\langle  Z_{ba}, \xi_{b a}\right\rangle}.
\end{split}
\ee
If inserting ${E_a^b}^i$ into the right hand side of Eq. \eqref{eq:brac}, the $P_{ab}$ becomes 
\be
\begin{split}
  {Q_a^b}^i&\equiv\left\langle \lambda j_{a b},-\vec{n}_{a b}\left| \left(E_{b}^{a}\right)^{i\dagger} Y^{\dagger} g_{a}^{-1} g_{b} Y\right| \lambda j_{a b}, \vec{n}_{b a}\right\rangle\\
  &\frac{d_{\lambda j_{a b}}}{\pi} \int_{\mathbb{C} \mathbb{P}^{1}} \mathrm{d} \tilde{\mathbf{z}}_{a b}\left\langle Z_{b a}, Z_{b a}\right\rangle^{-(1-\ii \gamma)\lambda j_{ab}}\left\langle Z_{a b}, Z_{a b}\right\rangle^{-(1+\ii \gamma )\lambda j_{a b}}\times\\
&\left\langle J \xi_{a b}, Z_{a b}\right\rangle^{2 \lambda j_{a b}}\left\langle Z_{b a}, \xi_{b a}\right\rangle^{2\lambda j_{a b}}(-\lambda j_{a b}\gamma)\frac{\left\langle J \xi_{a b}, \sigma_{i} Z_{ab}\right\rangle}{\left\langle J \xi_{a b}, Z_{ab}\right\rangle}
\end{split}
\ee
Defining  
\be\label{eq:ADE}
A^i_{ab}=\gamma \lambda j_{a b} \frac{\left\langle\sigma^{i} Z_{b a}, \xi_{b a}\right\rangle}{\left\langle Z_{b a}, \xi_{b a}\right\rangle},\ 
A^i_{ba}=-\gamma \lambda j_{ab}\frac{\left\langle J \xi_{ba}, \sigma_{i} Z_{ba}\right\rangle}{\left\langle J \xi_{ba}, Z_{ba}\right\rangle},
\ee
we have
\be
  {Q_a^b}^i=P_{ab}A^i_{ab},\ {Q^a_b}^i=P_{ab}A^i_{ba}.
\ee
Then we obtain the following integral expressions of the ingredients in \eqref{eq:prop1}
\begin{align}
  \bra{W}E^a_n\cdot E^b_n E^c_m \cdot E^d_m |\Psi_0\rangle&= \sum_{ j_{ab}}\psi_{\lambda j_0,\zeta_0}\int \mathrm{d} \phi\,  U(j,\phi) \, \lt[A_{an}(j,\phi) \cdot A_{bn}(j,\phi)\rt] \lt[A_{cm}(j,\phi) \cdot A_{dm}(j,\phi)\rt]\, \e^{\lambda S(j,\phi)},\label{eq:core1}\\
  \bra{W}E^a_n\cdot E^b_n |\Psi_0\rangle&=\sum_{ j_{ab}}\psi_{\lambda j_0,\zeta_0}\int \mathrm{d} \phi\,  U(j,\phi)  \,A_{an}(j,\phi) \cdot A_{bn}(j,\phi)\,\e^{\lambda S(j,\phi)},\label{eq:core2}\\
  \bra{W} E^c_m \cdot E^d_m |\Psi_0\rangle& = \sum_{j_{ab}}\psi_{\lambda j_0,\zeta_0}\int \mathrm{d} \phi \, U(j,\phi)\, A_{cm}(j,\phi) \cdot A_{dm}(j,\phi)\,\e^{\lambda S(j,\phi)}\label{eq:core3}.
\end{align}


\subsection{Large Spin Approximation}\label{Large Spin Approximation}
Equations \eqref{eq:core1}, \eqref{eq:core2}, and \eqref{eq:core3} are all expressed as summations of the $J_{ab}\equiv \lambda j_{ab}$ in the domain of non-negative half integers. This type of the summation can be rewritten as
\be\label{eq:prs}
\sum_{J\in \frac{\mathbb{Z}^+}{2} \cup 0} f(J) = \half \sum_{J\in\mathbb{Z}} f(|J/2|) +\half f(0)= 2\sum_{k\in\mathbb{Z}}\int_{0}^{\infty} dJ f(J) \e^{4\pi \ii k J}  +\half f(0),
\ee 
where use of the Poisson summation formula is made in the second step. Applying this formula to \eqref{eq:core1} - \eqref{eq:core3}, the term $\half f(0)$ is exponentially small when all the $\lambda j_0{}_{ab}$ are large in $\psi_{\lambda j_{0}, \phi_{0}}$ because the term contains a Gaussian peaked at $j_{ab}=j_0{}_{ab}$. By neglecting $\half f(0)$, \eqref{eq:core1} - \eqref{eq:core3} are expressed as integrals  
\begin{align}
  \langle W|\Psi_0\rangle&= (2\lambda)^{10}\sum_{\{k_{ab}\}\in\mathbb{Z}^{10}} \int_0^{\infty} \mathrm{d}^{10} j \int\mathrm{d} \phi \,U  \,\e^{-\lambda S_{tot}^{(k)} },\label{eq:core40}\\
  \bra{W}E^a_n\cdot E^b_n E^c_m \cdot E^d_m |\Psi_0\rangle&= (2\lambda)^{10}\sum_{\{k_{ab}\}\in\mathbb{Z}^{10}}\int_0^{\infty} \mathrm{d}^{10} j \int \mathrm{d} \phi\,  U \, \e^{-\lambda S_{tot}^{(k)} } (A_{an} \cdot A_{bn}) (A_{cm} \cdot A_{dm}),\label{eq:core50}\\
  \bra{W}E^a_n\cdot E^b_n |\Psi_0\rangle&=(2\lambda)^{10} \sum_{\{k_{ab}\}\in\mathbb{Z}^{10}}\int_0^{\infty} \mathrm{d}^{10} j \int \mathrm{d} \phi\, U\, \e^{-\lambda S_{tot}^{(k)} } (A_{an}  \cdot A_{bn} ),\label{eq:core60}\\
  \bra{W} E^c_m \cdot E^d_m |\Psi_0\rangle& =(2\lambda)^{10} \sum_{\{k_{ab}\}\in\mathbb{Z}^{10}}\int_0^{\infty} \mathrm{d}^{10} j \int \mathrm{d} \phi\,  U \, \e^{-\lambda S_{tot}^{(k)}}  (A_{cm}  \cdot A_{dm} )\label{eq:core70}, 
\end{align}
with the total action $S_{tot}^{(k)}$ written as
\begin{eqnarray}\label{eq:tol}
S_{tot}^{(k)}&=&S_{tot}+4\pi \ii\sum_{a>b}j_{ab}k_{ab},\\
S_{tot}&=& \ii \sum_{ab} \zeta_0^{ab}( j_{ab}-{j_{0}}_{ab})
+\sum_{ab,cd}\alpha^{(ab)(cd)} \frac{ j_{ab}-{ j_{0}}_{ab}}{\sqrt{{ j_{0}}_{ab}}}\frac{ j_{cd}-{ j_{0}}_{cd}}{\sqrt{{ j_{0}}_{cd}}} - S(j,\phi),
\end{eqnarray}
where $S$ is given in \eqref{eq:action1}. Recall that when $\lambda j_{ab}\in \Z/2$, $\lambda S$ is defined up to $2\pi i\Z$ because it contains logarithms which are multi-valued, so $e^{\lambda S}$ is single-valued. Nonetheless, when replacing the sums over $j_{ab}$ by the integral $\int\dd^{10}j$, $e^{\lambda S}$ becomes multi-valued since $j_{ab}$ becomes continuous. The integrands in (\ref{eq:core40}) - (\ref{eq:core70}) are understood as being defined on the covering space of the logarithms in $S$, while the integration domain of $\int\dd \phi$ is in the principle branch ${\rm Im}(\log(x))\in (-\pi,\pi]$ of the covering space. 

The critical point of $S_{tot}^{(k)}$ satisfies the following equations 
\be
\operatorname{Re}\left(S_{tot}\right)=0, \quad \partial_{j_{a b}} S_{tot}=-4 \pi \ii k_{a b}, \quad \partial_{\phi} S=0.\label{3crieqn}
\ee
The real part of the second equation above implies $j_{ab}=j_0{}_{ab}$ at the critical point ($\alpha$ has a positive definite real part). Then, $\operatorname{Re}\left(S_{tot}\right)=\operatorname{Re}\left(S\right)=0$ and $\partial_{\phi} S=0$ are the standard critical equations extensively studied in the asymptotics of 4-simplex amplitude (see e.g. \cite{Barrett:2009mw,HZ}). Given the boundary data in Section \ref{lab:bs}, there are 2 solutions (up to gauge freedom) of $\operatorname{Re}\left(S\right)=0$ and $ \partial_{\phi} S=0$ in the integration domain, corresponding to the Lorentzian 4-simplex geometry with opposite orientations. The 4-simplex geometry is consistent with the boundary data in Section \ref{lab:bs}. We denote these two solutions by $(j_0,\phi_0)$ and $(j_0,\phi'_0)$.

In order that Eqs.(\ref{3crieqn}) have solution. We demand the $\zeta_0^{ab}$ in Table \ref{tab:zeta} to satisfy the following relation
\be
i \zeta_{0}^{a b}=\frac{\partial {S}\left({j},\phi\right)}{\partial j_{a b}}\Big|_{{j}_0, \phi_0}\label{zeta0}.
\ee 
Here, $ \zeta_{0}^{a b}$ is defined modulo $4\pi\Z$ in $\psi_{\lambda j_0,\zeta_0}$ when $\lambda j_{ab},\lambda j_0{}_{ab}\in \Z/2$. When we replace the sums over $j_{ab}$ by integrals over continuous $j_{ab}$, this $4\pi\Z$ gauge symmetry is broken in each integral, so we fix the values of $\zeta_{0}^{a b}$ as in Table \ref{tab:zeta}.

The $\zeta_0^{ab}$ satisfying (\ref{zeta0}) gives $\partial_{j_{a b}} S_{tot}=0$ at one solution $(j_0,\phi_0)$ and $\partial_{j_{a b}} S_{tot}\neq0$ but $\partial_{j_{a b}} S_{tot}\not\in4\pi\mathbb{Z}$ at the other solution $(j_0,\phi'_0)$ \footnote{The spinfoam action $S=i\sum_{a>b}j_{ab}(\pm\gamma\Theta_{ab}+2\theta_{ab})$ at 2 solution \cite{Barrett:2009mw}. $\Theta_{ab}$ are the dihedral angles of the 4-simplex, and $\theta_{ab}=\psi_{ab}-\psi_{ba}$ give the overall phase of the asymptotics. When we set $\zeta_0^{ab}=\gamma\Theta_{ab}+2\theta_{ab}$, we have $\partial_{j_{a b}} S_{tot}=0$ at one solution and $\partial_{j_{a b}} S_{tot}=2\gamma\Theta_{ab}\not\in 4\pi\mathbb{Z}$ at the other solution.}. Therefore, any $k_{ab}\neq 0$ leads to Eqs.(\ref{3crieqn}) have no solution. When $\lambda$ is large, the integrals in \eqref{eq:core40} - \eqref{eq:core70} are suppressed exponentially, unless all $k_{ab}=0$. When $k_{ab}=0$, Eqs.\eqref{3crieqn} have a unique solution $(j_0,\phi_0)$. which is the critical point of the integral.

Furthermore, because of the Gaussian in $\psi_{\lambda j_0,\zeta_0}$, the integrals \eqref{eq:core40} - \eqref{eq:core70} are dominant in the neighborhood where $j_{ab}$ is close to the $j_0{}_{ab}$. When $\lambda$ is large, we can approximate the $\int_0^{\infty}$ in these integrals to $\int_{-\infty}^{\infty}$, while their differences are exponentially suppressed.

We rewrite \eqref{eq:core40} - \eqref{eq:core70} as below
\begin{align}
  \langle W|\Psi_0\rangle&\simeq  \int_{-\infty}^{\infty} \mathrm{d}^{10} j \int\mathrm{d} \phi \,\tilde{U}  \,\e^{-\lambda S_{tot} },\label{eq:core4}\\
  \bra{W}E^a_n\cdot E^b_n E^c_m \cdot E^d_m |\Psi_0\rangle&\simeq \int_{-\infty}^{\infty} \mathrm{d}^{10} j \int \mathrm{d} \phi\,  \tilde{U} \, \e^{-\lambda S_{tot} } (A_{an} \cdot A_{bn}) (A_{cm} \cdot A_{dm}),\label{eq:core5}\\
  \bra{W}E^a_n\cdot E^b_n |\Psi_0\rangle&\simeq \int_{-\infty}^{\infty} \mathrm{d}^{10} j \int \mathrm{d} \phi\, \tilde{U}\, \e^{-\lambda S_{tot} } (A_{an}  \cdot A_{bn} ),\label{eq:core6}\\
  \bra{W} E^c_m \cdot E^d_m |\Psi_0\rangle& \simeq \int_{-\infty}^{\infty} \mathrm{d}^{10} j \int \mathrm{d} \phi\,  \tilde{U} \, \e^{-\lambda S_{tot}}  (A_{cm}  \cdot A_{dm} )\label{eq:core7}, 
\end{align}
where $\tilde{U}\equiv(2\lambda)^{10} U$. Our numerical program computes expectation values 
\be
\frac{\bra{W}E^a_n\cdot E^b_n E^c_m \cdot E^d_m |\Psi_0\rangle}{\langle W|\Psi_0\rangle},\quad \frac{\bra{W}E^a_n\cdot E^b_n |\Psi_0\rangle}{\langle W|\Psi_0\rangle},\quad \frac{\bra{W} E^c_m \cdot E^d_m |\Psi_0\rangle}{\langle W|\Psi_0\rangle}\label{expvalues}
\ee
by their approximate integral expressions in \eqref{eq:core4} - \eqref{eq:core7}. Comparing to the original definition in (\ref{eq:prop1}), we have neglected 3 types contributions, $\half f(0)$ in (\ref{eq:prs}), $k_{ab}\neq 0$, and $\int_{-\infty}^{0}\mathrm{d} j_{ab}$, that are exponentially suppressed when $\lambda$ is large.

In order to apply the method of Lefschetz thimble, in (\ref{eq:core4}) - (\ref{eq:core7}), we extend the integration domain of $\int \dd\phi$ on the cover space beyond the principle branch to allow $\mathrm{Im}(\log(x))\in (-\infty,\infty)$, so that $\int \dd\phi$ is along the integration cycle connecting to the infinity of the cover space. (\ref{eq:core4}) - (\ref{eq:core7}) have no critical point beyond the principle branch (by fixing the values of $\zeta_0^{ab}$). Therefore, extending the integrals only add contributions that are exponentially suppressed at large $\lambda$.

\subsection{Critical Point}\label{subsec:critical}
Finding the critical point of the total action $S_{tot}$ is important for both the asymptotic expansion method and our Lefschetz thimble method. The action $S_{tot}$ has only one critical point corresponding to the aforementioned 4-simplex geometry. As the solution to the critical equations ${\rm Re}(S_{tot})=0$, $\partial_\phi S_{tot}=0$, and $\partial_j S_{tot}=0$, the critical values of the group elements ${g_{0}}_{a},\ (a=2\cdots5)$ are the spinor representations of the Lorentz transformations converting $N_2\cdots N_5$ in \eqref{eq:4Normals} to $(1,0,0,0)$ \footnote{More details are found \cite{Dona:2019dkf,Han:2020fil}}. TABLE \ref{tab:ga} lists the explicit values of ${g_{0}}_{a},\ (a=2\cdots5)$. 
\begin{table}[h]
	\centering\caption{Each cell of the table is the critical point of a-th group element.}\label{tab:ga}
	\scriptsize
	\setlength{\tabcolsep}{0.5mm}
	\begin{tabular}{|c|c|c|c|c|c|}
		\hline
		\small{a}&1&2&3&4&5\\
		\hline
		${g_{0}}{}_{a}$ &$\left(\begin{matrix}
		1&0\\	0&1
		\end{matrix}\right)$&$\left(\begin{matrix}
		0.18 \ii&1.01 \ii\\	1.01 \ii&0.18 \ii
		\end{matrix}\right)$&$\left(\begin{matrix}
		0.18 \ii&0.96 - 0.34 \ii\\	-0.96- 0.34 \ii&0.18 \ii
		\end{matrix}\right)$&$\left(\begin{matrix}
		1.01 \ii&-0.48 - 0.34 \ii\\	0.48 - 0.34 \ii&-0.65 \ii
		\end{matrix}\right)$&$\left(\begin{matrix}
		-0.65 \ii&-0.48 - 0.34 \ii\\	0.48 - 0.34 \ii&1.01 \ii
		\end{matrix}\right)$\\
		\hline
	\end{tabular}
\end{table} 

Having had the critical values of the group elements, we can plug them into the critical equations ${\rm Re}(S_{tot})=0$, which is equivalent to
\be
\ket{\xi_{ab}}=\frac{e^{i\psi _{ab}}}{\left\Vert Z_{ab}\right\Vert }g^{\dagger}_{a}\ket{z_{ab}},\quad
\text{and}\quad \ket{J\xi_{ba}}=\frac{e^{i\psi _{ba}}}{\left\Vert Z_{ba}\right\Vert }g^{\dagger}_{b}\ket{z_{ab}},  \label{eq:ReS}
\ee
with  $\left\Vert Z_{ab}\right\Vert = \left\vert \left\langle
Z_{ab},Z_{ab}\right\rangle \right\vert ^{1/2}$, in order to determine the spinors $z_{ab}$ and the phase factors $\psi_{ba}$, $\psi_{ab}$. The normalized components of the spinors $z_{ab}$ are recorded in TABLE \ref{tab:zv}.
\begin{table}[htbp]
  \centering\caption{Each cell indicates a spinor $z_0{}_{ab}$.}\label{tab:zv}
\footnotesize
\setlength{\tabcolsep}{0.8mm}
\begin{tabular}{|c|c|c|c|c|c|}
  \hline
  \diagbox{\small{a}}{$\ket{z_0{}_{ab}}$ }{\small{b}}&1&2&3&4&5\\
  \hline
  1&\diagbox{}{}&(1,1)&(1,-0.333+0.942i)&(1,-0.184-0.259i)&(1,-1.817-2.569 i)\\
  \hline
  2&(1,1)&\diagbox{}{}&(1,0.685-0.729i)&(1, 1.857 + 0.989 i)&(1, 0.420 + 0.223 i)\\
  \hline
  3&(1, 0.333 - 0.943 i)&(1, 0.685 - 0.729 i)&\diagbox{}{}&(1, 0.313 + 2.080 i)&(1, 0.071 + 0.470 i)\\
  \hline
  4&(1, -0.184-0.259i)&(1, 1.857 + 0.989 i)&(1,0.313 + 2.080 i)&\diagbox{}{}&(1, 0.058+0.082i)\\
  \hline
  5&(1, -1.817-2.569 i)&(1,  0.420 + 0.223 i)&(1,0.071 + 0.470 i)&(1,  0.058+0.082i)&\diagbox{}{}\\
  \hline
\end{tabular}
\end{table}

Based on the critical $g_0{}_{ab}$ and $z_0{}_{ab}$, we apply our parameterizations \eqref{eq:par1} and \eqref{eq:par2} but let them centered at the critical point:
\be\label{eq:par3}
g=g_0\left(
\begin{matrix}
  1+\frac{x_1+\ii y_1}{\sqrt{2}} && \frac{x_2+\ii y_2}{\sqrt{2}}\\
  \frac{x_3+\ii y_3}{\sqrt{2}} && \frac{1+\frac{x_2+\ii y_2}{\sqrt{2}}\frac{x_3+\ii y_3}{\sqrt{2}}}{1+\frac{x_1+\ii y_1}{\sqrt{2}}}\\
\end{matrix}
\right)
,\quad
\text{and}\quad 
z=(1,z_0+x+\ii y), 
\ee
where $g_0$ stands for the critical value of the group variable in TABLE \ref{tab:ga} and $z_0$ is the second component of the spinor variable in TABLE \ref{tab:zv}. When all the parameters are zero, the group variables and spinors take their critical values.

To conclude, here comes 2 main points of this section: Firstly, the total action $S_{tot}$ in \eqref{eq:tol} depends on $54$ real parameters, and specifically we parametrize the four group variables $g_a$ and the ten spinor variables $z_{ab}$ as \eqref{eq:par3}. Secondly, the spinfoam propagator $G^{abcd}_{mn}$ can be computed by the integrals \eqref{eq:core4} - \eqref{eq:core7}. In the next section, we will show the algorithm to evaluate these integrals. 

\section{Theory and Algorithm}\label{sec:TM}

The total action $S_{tot}$ is a complex valued, such that the integrands in \eqref{eq:core4} - \eqref{eq:core7} are highly oscillatory, especially when $\lambda$ is large. This fact plagues the attempts of using the conventional Monte-Carlo method to compute the spinfoam propagator. In this section, we review how to use \textit{Picard-Lefschetz theory} to transform these types of integrals to be non-oscillatory (see e.g. \cite{Witten:2010cx,Alexandru:2020wrj}) for reviews), and we present the algorithm that combines the thimble and Markov-Chain Monte Carlo (MCMC) methods and can compute the expectation value of an observable when the action is complex valued. 

\subsection{Lefschetz Thimble}\label{subsec:thimbleframework}
The thimble method is a high-dimensional generalization the elementary saddle point integration along the steepest decent (SD). The thimble method can be further generalized as follows.

  The starting point of computing the integral
  \be\label{eq:AA}
  A=\int \dd^n x f(\vec{x}) \mathrm{e}^{-S(\vec{x})},
  \ee
  is to analytically continue both $f(\vec{x})$ and $S(\vec{x})$ to be holomorphic functions $\hat{f}(\vec{z})$ and $\hat{S}(\vec{z})$, such that \eqref{eq:AA} becomes an integral of analytic functions $\hat{f}(\vec{z})$ and $\hat{S}(\vec{z})$ of complex variables on the real domain
  \be
  A=\int_{\mathbb{R}^n} \dd^n z \hat{f}(\vec{z}) \mathrm{e}^{-\hat{S}(\vec{z})},
  \ee 
 where $\dd^n z \hat{f}(\vec{z}) \mathrm{e}^{-\hat{S}(\vec{z})}$ is a holomorphic $n$-form restricted on $\R^n$.
  
The Picard-Lefschetz theory shows that the integral $A$ can be decomposed into a linear combination of integrals over real $n$-dimensional integral cycles $\mathcal{J}_\sigma,\ \sigma=1,2,3,\cdots$
  \be\label{eq:decomT}
    \int_{\mathbb{R}^n} \dd^n z \hat{f}(\vec{z}) \mathrm{e}^{-\hat{S}(\vec{z})}=\sum_{\sigma} n_\sigma \int_{\mathcal{J}_\sigma} \dd^n z \hat{f}(\vec{z}) \mathrm{e}^{-\hat{S}(\vec{z})},
  \ee
when $\mathbb{R}^n$ is homologically equivalent to $\sum_\sigma n_\sigma \mathcal{J}_\sigma$. The holomorphic $n$-form $\dd^n z \hat{f}(\vec{z}) \mathrm{e}^{-\hat{S}(\vec{z})}$ is restricted in a class of ${\mathcal{J}_\sigma}$ on the right hand side. This decomposition is given by the $n$-dimensional real sub-manifolds $\mathcal{J}_{\sigma}$, each attached to a critical point $p_{\sigma}$ satisfying $\partial_z\hat{S}(p_{\sigma})=0$ \footnote{We do not impose $\mathrm{Re}(\hat{S})=0$ for critical points in the complex space.}. Each $\mathcal{J}_{\sigma}$, called a Lefschetz thimble, is a union of SD paths that are solutions to the SD equations 
  \begin{equation}\label{eq:SDeq}
      \frac{\dd z^a }{ \dd t}=-\frac{\partial\overline{\hat{S}(\vec{z})}}{\partial\overline{z^a}},
  \end{equation}
and falls to the critical point $p_\sigma$ when $t\to\infty$. Here we call $t$ the flow time. 

Since 
  \be
  \frac{\dd \hat{S}}{\dd t}=\frac{\partial \hat{S}}{\partial z^a} \frac{\dd z^a}{\dd t}=-\left\vert\frac{\partial \hat{S} }{\partial z^a} \right\vert^2,
  \ee
$\mathrm{Re}(\hat{S})$ monotonically decreases along each SD path and approaches its minimum at the critical point, while $\mathrm{Im}(\hat{S})$ is conserved along each path. Thus, on each thimble $\mathcal{J}_{\sigma}$, 
  \be\label{eq:integralonThimble}
  \int_{\mathcal{J}_\sigma} \dd^n z \hat{f}(\vec{z}) \mathrm{e}^{-\hat{S}(\vec{z})}=\mathrm{e}^{-\ii\, {\rm Im}{(\hat{S}(p_{\sigma}))}}\int_{\mathcal{J}_\sigma} \dd^n z \hat{f}(\vec{z}) \mathrm{e}^{-{\rm Re}{(\hat{S}(\vec{z}))}}
  \ee
  becomes a non-oscillatory integral times a constant phase $\mathrm{e}^{-\ii\, {\rm Im}{(\hat{S}(p_{\sigma}))}}$. On $\mathcal{J}_\sigma$, ${\rm Re}(\hat{S})$ grows when moving far away from the critical point, so the integrand is exponentially suppressed at the infinity, and the integral on $\mathcal{J}_\sigma$ is convergent. 
  
  The Lefschetz-Thimbles $\{ \mathcal{J}_{\sigma} \}$ presents a good basis of relative homology group for the integral \eqref{eq:integralonThimble} \cite{Scorzato:2015qts}. Using this basis, the integral \eqref{eq:decomT} is valid for a specific set $\{ n_\sigma \}$ of the weights of the thimbles. Consider $\hat{f}$ as an observable. The expectation value $\langle f \rangle$ is given by
  \be\label{eq:obs1}
  \langle f \rangle
  =\frac{\int_{\mathbb{R}^n} \dd^n z \hat{f}(\vec{z}) \mathrm{e}^{-\hat{S}(\vec{z})}}{\int_{\mathbb{R}^n} \dd^n z  \mathrm{e}^{-\hat{S}(\vec{z})}}
  =\frac{\sum_{\sigma} n_\sigma \int_{\mathcal{J}_\sigma} \dd^n z \hat{f}(\vec{z}) \mathrm{e}^{-\hat{S}(\vec{z})}}{\sum_{\sigma} n_\sigma \int_{\mathcal{J}_\sigma} \dd^n z  \mathrm{e}^{-\hat{S}(\vec{z})}}.
  \ee
  A weight $n_\sigma$ is the intersection number between the original integration cycle $\mathbb{R}^n$ and the manifold of the steepest ascent (SA) paths approaching the critical point $p_\sigma$ as $t\to\infty$. The SA paths are solutions to the SA equations
  \be\label{eq:SA}
    \frac{\dd z^a }{ \dd t}=\frac{\partial\overline{\hat{S}(\vec{z})}}{\partial\overline{z^a}}.
  \ee
Along each SA path, $\mathrm{Re}(\hat{S})$ monotonically increases and approach the maximum at the critical point, while $\mathrm{Re}(\hat{S})$ is conserved along the path. Computing these weights are challenging in general (see e.g. \cite{Bedaque:2017epw,Bluecher:2018sgj} for some recent progresses). Nevertheless, in the cases where one of the thimble, denoted by $\mathcal{J}_{\sigma'}$, dominates the integral, we may neglect the contribution of other thimbles and re-express \eqref{eq:obs1} as
  \be\label{eq:regThimble}
  \langle f \rangle
  \simeq \frac{n_{\sigma'} \mathrm{e}^{-\mathrm{i}\, {\rm Im}(S(p_{\sigma'}))} \int_{\mathcal{J}_{\sigma'}} \dd^n z\, \hat{f}(\vec{z}) \mathrm{e}^{-{\rm Re} (\hat{S}(\vec{z}))}}{ n_{\sigma'} \mathrm{e}^{-\mathrm{i}\, {\rm Im}(S(p_{\sigma'}))} \int_{\mathcal{J}_{\sigma'}} \dd^n z\,  \mathrm{e}^{-{\rm Re}(\hat{S}(\vec{z}))}}
  =\frac{ \int_{\mathcal{J}_{\sigma'}} \dd^n z\, \hat{f}(\vec{z}) \mathrm{e}^{-{\rm Re} (\hat{S}(\vec{z}))}}{ \int_{\mathcal{J}_{\sigma'}} \dd^n z \, \mathrm{e}^{-{\rm Re}(\hat{S}(\vec{z}))}},
  \ee  
  which can be considered as a mean value provided by a sampling on the thimble $\mathcal{J}_{\sigma'}$ with a Boltzmann factor $\e^{-{\rm Re}(\hat{S}(\vec{z}))}$. Then, it is possible to use the MCMC method to numerically compute $\langle f \rangle$. Each integral involved in the spinfoam propagator has a single critical point in its integration domain. The Lefschetz thimble of the critical point is dominant. Thus the computation of the spinfoam propagator is the case where Eq. \eqref{eq:regThimble} applies.

\subsection{Thimbles Generated by Flows}\label{ssec:TGF}

\begin{figure}[h]
  \centering\includegraphics[width=0.95\textwidth]{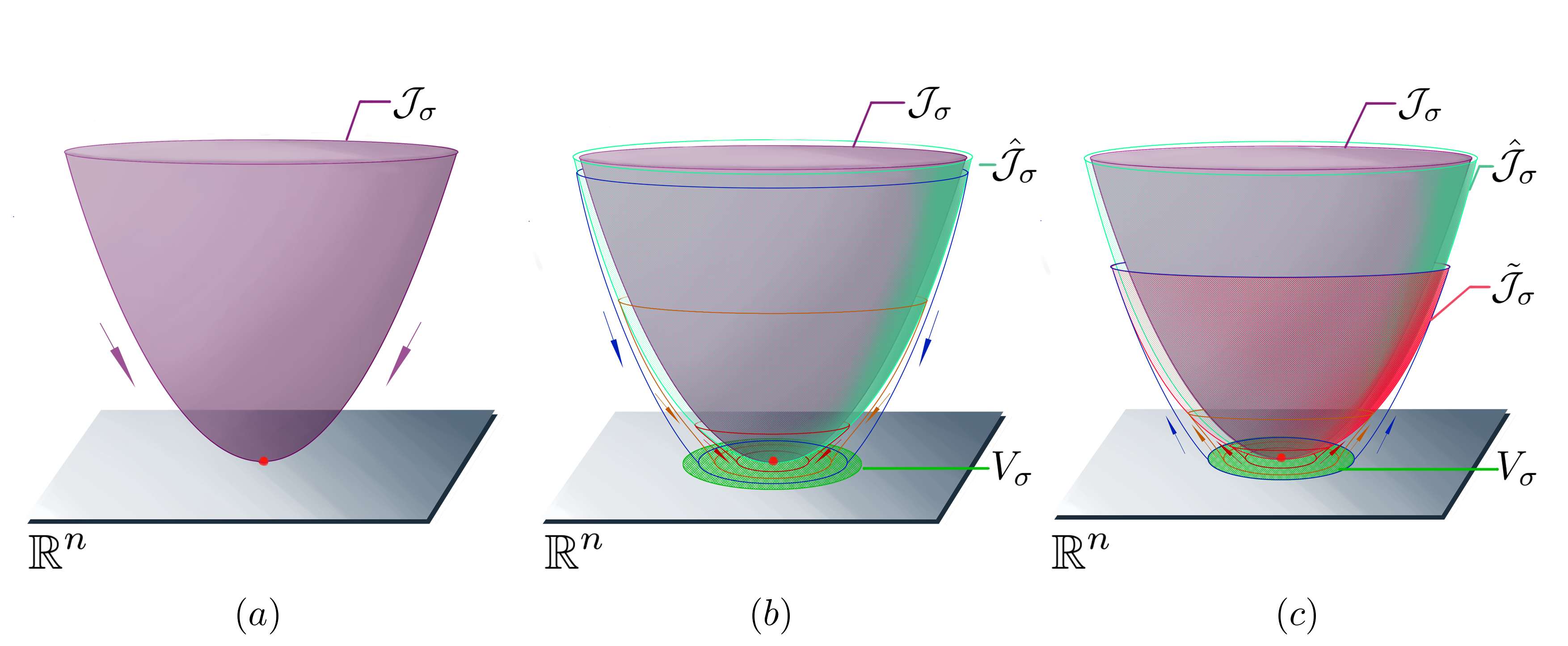}
  \caption{
  (a) A Lefschetz thimble $\mathcal{J}_{\sigma}$ (purple surface) is the union of all the SD paths falling to the critical point $p_\sigma$ (red dot) when $t\to\infty$. \\
  (b) $\widehat{\mathcal{J}}_\sigma$ (green transparent surface) is defined as the union of points that can flow to $V_\sigma$ (green disk at the bottom) after an infinite time evolution by the SD equation. For example, the cross-sections of $\widehat{\mathcal{J}}_\sigma$ (illustrated by the blue, yellow and red circles in $\widehat{\mathcal{J}}_\sigma$) flow to the cross-sections in $V_\sigma$ (blue, yellow and red circles in the green disk). \\
  (c) $\tilde{\mathcal{J}}_\sigma$ (red transparent surface) is generated by upward flowing every points in $V_\sigma$ with a finite time evolution by the SA equation. The cross-sections in $V_\sigma$ (blue, yellow and red circles in the green disk) flow upward to the cross-sections in $\tilde{\mathcal{J}}_\sigma$ (blue, yellow and red circles in $\tilde{\mathcal{J}}_\sigma$).
  }\label{fig:thimble1}
\end{figure}

As the first step to apply the the Lefschetz thimble framework to numerics, we need to find the Lefschetz thimble $\mathcal{J}_\sigma$ for a given critical point $p_\sigma$ (FIG. \ref{fig:thimble1} (a)). By definition, one might try to decide if a point is on the thimble by checking if it falls to $p_\sigma$ after flowing infinitely long time described by the SD equation; however, it is hard in practice due to the infinite flow time. Naively, we might also use $p_\sigma$ as the initial point and using SA equation to generate the thimble as the union of the paths going away from the $p_\sigma$, but this way is also problematic because $p_\sigma$ is a fixed point of the SA equation. 

We follow the method reviewed in \cite{Alexandru:2020wrj} to bypass the difficulty of generating $\mathcal{J}_{\sigma}$ numerically. We consider a small real $n$-dimensional neighborhood $V_{\sigma}$ of the critical point $p_\sigma$ and a slightly different integral cycle denoted by $\widehat{\mathcal{J}}_{\sigma}$. $\widehat{\mathcal{J}}_{\sigma}$ is the union of solutions to the SD equations \eqref{eq:SDeq} flowing to $V_{\sigma}$ after infinite time evolution. $\widehat{\mathcal{J}}_{\sigma}$ is also real $n$- dimensional. $\widehat{\mathcal{J}}_{\sigma}$ is a good approximation of the true thimble $\mathcal{J}_{\sigma}$ when the size of $V_{\sigma}$ is small, as all the SD paths on $\widehat{\mathcal{J}}_{\sigma}$ connect with $V_{\sigma}$. $\widehat{\mathcal{J}}_{\sigma}$ (FIG. \ref{fig:thimble1} (b)) approaches $\mathcal{J}_{\sigma}$ when $V_{\sigma}$ shrinks to the critical point $p_\sigma$. Since the integrand is analytic, and $\widehat{\mathcal{J}}_{\sigma}$ is a deformation of $\mathcal{J}_{\sigma}$, the integral 
\be\label{eq:appo1}
\int_{\widehat{\mathcal{J}}_{\sigma}} d^n z\hat{f}(\vec{z})e^{-\hat{S}(\vec{z})}
\ee
is the same as \eqref{eq:integralonThimble}; however, since ${\rm Im}{(\hat{S})}$ is no longer constant in $\widehat{\mathcal{J}}_{\sigma}$, the above integral becomes oscillatory in contrast to the integral on $\mathcal{J}_{\sigma}$. If $V_{\sigma}$ is small enough, however, the fluctuation of the ${\rm Im}{(\hat{S})}$ on $\widehat{\mathcal{J}}_{\sigma}$ is so small that the oscillation of the integral is weak enough to keep the Monte Carlo method accurate.


Since infinite time evolution is involved, finding the entire $\widehat{\mathcal{J}}_{\sigma}$ is not numerically practical. A practical integral cycle $\tilde{\mathcal{J}}_{\sigma}$ is the union of the solutions to the SD equations \eqref{eq:SDeq} falling to $V_{\sigma}$ after finite but sufficiently long flow time. The thimble $\tilde{\mathcal{J}}_{\sigma}$ approaches $\widehat{\mathcal{J}}_{\sigma}$ when the flow time is infinite. Similar to the method in \cite{Bedaque:2017epw}, we can find the approximate $\tilde{\mathcal{J}}_{\sigma}$ in an inverse process. Namely, we choose a small real $n$-dimensional neighborhood $V_{\sigma}$ of the critical point, and flow upward from points in $V_{\sigma}$ according to the SA equation with a finite time $T$. The end points of these flows form a real $n$-dimensional manifold $\tilde{\mathcal{J}}_{\sigma}$ (FIG. \ref{fig:thimble1}). The thimble $\tilde{\mathcal{J}}_{\sigma}$ does not reach the infinity of the Lefschetz thimble $\mathcal{J}_\sigma$. The size of $\tilde{\mathcal{J}}_{\sigma}$ depends on the choice of $T$. 

Our two-step approximation of $\mathcal{J}_{\sigma}$ is illustrated in the following diagram: 
\[
\mathcal{J}_\sigma\xrightarrow{\text{Fix } V_{\sigma}}\widehat{\mathcal{J}}_\sigma\xrightarrow{\text{Fix } T}\tilde{\mathcal{J}}_\sigma.
\]
In the first step, we use the $\widehat{\mathcal{J}}_\sigma$ as the union of all the steepest decent paths falling to $V_{\sigma}$ to approximate $\mathcal{J}_{\sigma}$. In our computation, we set the size of $V_{\sigma}$ by setting a tolerance of the fluctuation of the ${\rm Im}{(\hat{S})}$ on $\widehat{\mathcal{J}}_\sigma$. In the second step, we use $\tilde{\mathcal{J}}_\sigma$ as the union of the finitely evolved steepest ascent paths starting from the points in $V_{\sigma}$ to approximate $\widehat{\mathcal{J}}_\sigma$. Thus, the longer $T$ and smaller $V_{\sigma}$ are, the better approximation of $\mathcal{J}_{\sigma}$ is achieved as $\widehat{\mathcal{J}}_\sigma$.

Another remark is that in the second step of the approximation, making $\tilde{\mathcal{J}}_\sigma$ very large is actually unnecessary. When computing \eqref{eq:appo1}, we sample the points on the thimble with the probability distribution $\e^{-{\rm Re}{(S)}}$, and the contributions to the integral from points far away from the critical point are exponentially suppressed. Thus, we can choose the $T$ parameter, which provides sufficiently large $\tilde{\mathcal{J}}_\sigma$ containing points that contribute dominantly to \eqref{eq:appo1}, while ensuring increasing $T$ only add negligible contribution to the integral. The result should converge when further increasing $T$. In fact, experiences from our computation and other existing results \cite{Alexandru:2020wrj,Alexandru:2015sua,Alexandru:2015xva} suggest even $T<1$ suffice to result in good accuracy.

The choice of real $n$-dimensional $V_{\sigma}$ depends on the local behavior of the SA equations \eqref{eq:SA} around the critical point $p_\sigma$. 
Consider a small holomorphic variation $\omega^k=\delta z^k$, we linearize \eqref{eq:SA}:
\be\label{eq:jacobian}
    \frac{\dd \omega^k}{\dd t}=\overline{\frac{\partial^2 \hat{S}}{\partial z_k \partial z_l }} \cdot \overline{\omega^l},
\ee
In the neighborhood of $p_\sigma$, $\overline{\frac{\partial^2 \hat{S}}{\partial z_k \partial z_l }}$ can be approximated by the Hessian $\bar{\mathbf{H}}$ of $\hat{S}(\vec{z})$ at $p_{\sigma}$, and solution of \eqref{eq:jacobian} is given by 
\[
{\omega}=\sum_{a=1}^{2n}\e^{\lambda^a t} {\omega}_a,
\]
where $\lambda^a$ and ${\omega}_a$ are the eigenvalues and the corresponding eigenvectors of the generalized eigenvalue equation: 
\be\label{eq:HES1}
\mathbf{H} {\omega} = \lambda \bar{{\omega}}.
\ee
Using the Takagi factorization \cite{192483}, we covert \eqref{eq:HES1} to a real $2n$-dimensional eigenvalue equation:
\be\label{eq:HES2}
\left[
\begin{matrix}
  \mathbf{H}_{\mathbb{R}}&&\mathbf{H}_{\mathbb{I}}\\
  \mathbf{H}_{\mathbb{I}}&&-\mathbf{H}_{\mathbb{R}}\\
\end{matrix}
\right]
\left[
\begin{matrix}
  {\omega}_{\mathbb{R}}\\
  {\omega}_{\mathbb{I}}\\
\end{matrix}
\right]
=\lambda \left[
  \begin{matrix}
    {\omega}_{\mathbb{R}}\\
   {\omega}_{\mathbb{I}}\\
  \end{matrix}
  \right],
\ee
where $\mathbf{H}_{\mathbb{R}}$ and $\mathbf{H}_{\mathbb{I}}$ are the real and imaginary parts of the Hessian $\mathbf{H}$. The eigenvalues $\lambda$ in \eqref{eq:HES2}, which are equivalent to $\lambda$ in \eqref{eq:HES1}, come in pairs $\{ \pm \lambda_i \}$. The eigenvectors $({\mathbb{\omega}_i}{}_{\mathbb{R}},{\mathbb{\omega}_i}{}_{\mathbb{I}})$, called the Takagi vectors, can reconstruct the eigenvectors $\{ {\omega}_i \}$ by
\[
 {\omega}_i={{\omega}_i}{}_{\mathbb{R}}+\ii {{\omega}_i}{}_{\mathbb{I}}.
\]
The flow given by \eqref{eq:SA} is repulsive along the eigenvectors $\mathbb{\omega}_a$ with positive eigenvalues, and is attractive along the eigenvectors $\mathbb{\omega}_a$ with negative eigenvalues. The paths along the attractive directions converge to the critical point $p_\sigma$, so they are not the paths that can form $\tilde{\mathcal{J}}_\sigma$. Only the paths flowing along the repulsive directions can form the $\tilde{\mathcal{J}}_\sigma$. We denote ${\hat{\omega}}_a$ as the normalized eigenvectors with positive eigenvalues. The space $\hat{V}_\sigma$ understood as a local neighborhood in the $\vec{z}$-coordinate chart at $p_\sigma$ is expressed as 
\be\label{eq:basis}
\hat{V}_\sigma=\{ \vec{z}|\vec{z}=\sum_{a=1}^n {\hat{\omega}}_i x^i+\vec{z}_\sigma,\ \text{each }{x}^i\in\mathbb{R}\text{ is small}\}.
\ee 
where $\vec{z}_\sigma$ are coordinates of $p_\sigma$. $\hat{V}_\sigma$ turns out to be the best choice of $V_{\sigma}$.

By $\delta z^k=(\partial z^k/\partial x^i)\delta x^i$ for the coordinates $\{x^i\}$ on $\hat{V}_\sigma$, and assuming $\dd \delta x^i/\dd t=0$, $J^k_i\equiv \partial z^k/\partial x^i$ satisfies the same equation as \eqref{eq:jacobian}, i.e.
\be\label{eq:jacobian1}
    \frac{\dd (J_{i}^{k})_t}{\dd t}=\sum_{l=1}^n\overline{\frac{\partial^2 \hat{S}}{\partial z_k \partial z_l }} (\overline{J_{i}^{l})_t}.
\ee
The solution $J_t$ is the Jacobian matrix of a flow of coordinate changes from $\{x^i\}$ to $\{z^i\}$. The initial condition $J_0$ is the constant $n\times n$-matrix, whose columns are the vectors $\hat{\omega}_a$. In what follows, $J:=J_T$ is the Jacobian for changing from $\{x^i\}$ to $\{z^i\}$ on $\tilde{\mathcal{J}}_\sigma$.

By the coordinate change, for any holomorphic function $\psi(z)$, its integral on $\tilde{\mathcal{J}}_\sigma$ can be expressed by the integral of $\{x^i\}$ in ${\hat{V}_\sigma}$
\be
\int_{\tilde{\mathcal{J}}_\sigma} \dd^n z\, \psi(z) =\int_{\hat{V}_\sigma}\dd^n x\, \det(J(x))\,\psi(z(x)).\label{pullback}
\ee


In the active point of view, for a fixed flow time $T$, every point in $\hat{V}_{\sigma}$ flows upward to $\tilde{\mathcal{J}}_\sigma$ according to the SA equation \eqref{eq:SA}. We define the local diffeomorphism
\[
\mathcal{C}_T:\hat{V}_{\sigma}\to\tilde{\mathcal{J}}_\sigma,
\]
that can map the initial point $p\in\hat{V}_{\sigma}$ to the end point $\mathcal{C}_T(p)\in\tilde{\mathcal{J}}_\sigma$ of the SA path with the finite evolution time $T$. The coordinate change from $\{x^i\}$ to $\{z^i\}$ is induced by $\mathcal{C}_T$.

As a result, for any given observable $f$, its expectation value can be computed by 
\be\label{eq:exc}
\begin{split}
\langle f \rangle& 
\simeq \frac{ \int_{\tilde{\mathcal{J}}_{\sigma}} \dd^n z \,\hat{f}(z)\, \mathrm{e}^{- \hat{S}(z)}}{ \int_{\tilde{\mathcal{J}}_{\sigma}} \dd^n z \, \mathrm{e}^{-\hat{S}(z)}}
= \frac{ \int_{\hat{V}_{\sigma}} \dd^n {x}\, \det(J(x))\, \hat{f}\lt(\mathcal{C}_T(x)\rt)\, \mathrm{e}^{- \hat{S}\lt(\mathcal{C}_T(x)\rt)}}{ \int_{\hat{V}_{\sigma}} \dd^n {x}\, \det(J(x))\,  \mathrm{e}^{-\hat{S}\lt(\mathcal{C}_T(x)\rt)}}\\
&= \frac{ 
  \int_{\hat{V}_{\sigma}} \dd^n {x}\, \e^{\ii(\arg(\det(J))-{\rm Im}(\hat{S}))}\, \hat{f}\, \mathrm{e}^{- {\rm Re}(\hat{S}) + \log(|\det(J)|)} 
}{ 
  \int_{\hat{V}_{\sigma}} \dd^n  {x}\, \e^{\ii(\arg(\det(J))-{\rm Im}(\hat{S}))}\, \mathrm{e}^{- {\rm Re}(\hat{S}) + \log(|\det(J)|)}
},\\
\end{split}
\ee
where in the second step we apply \eqref{pullback}. Note that $\det(J)$ is a complex number, and that $\log(\det(J))$ is given by $\log(|\det(J)|)+\ii \arg(\det(J))$.

We define ${\rm Re}(\hat{S}) - \log(\det(J))\equiv S_{eff}$ as the purely real effective action. For any observable $\mathcal{O}$, we define its expectation with respect to the effective action as 
\be
\langle \mathcal{O} \rangle_{eff}=\frac{ \int_{\hat{V}_{\sigma}} \dd^n {x} \,\mathcal{O}\, \mathrm{e}^{- S_{eff}} }{ \int_{\hat{V}_{\sigma}} \dd^n {x}\, \mathrm{e}^{- S_{eff}}}.
\ee

We define $\arg(\det(J))-{\rm Im}(\hat{S})$ as the residual phase $\theta_{res}$, and rewrite \eqref{eq:exc} as
\be\label{eq:effs}
\langle f \rangle \simeq \frac{ 
  \int_{\hat{V}_{\sigma}} \dd^n {x}\, \hat{f}\,\e^{\ii \theta_{res}}\,  \mathrm{e}^{- S_{eff}}
}{ 
  \int_{\hat{V}_{\sigma}} \dd^n {x}\,  \mathrm{e}^{-S_{eff}}
} \times\frac{ 
  \int_{\hat{V}_{\sigma}} \dd^n {x}\,  \mathrm{e}^{-S_{eff}} 
}{ 
  \int_{\hat{V}_{\sigma}} \dd^n {x} \,\e^{\ii \theta_{res}} \,\mathrm{e}^{- S_{eff}}
}
=\frac{\langle \e^{\ii\theta_{res}} \hat{f} \rangle_{eff}}{\langle \e^{\ii\theta_{res}} \rangle_{eff}}.
\ee
Note that ${\rm Im}(\hat{S}(\vec{z}))$ is not a constant in $\hat{V}_{\sigma}$. The fluctuation of $\mathrm{Im}(\hat{S}(\vec{z}))$ tends to vanish when shrinking $\hat{V}_\sigma$. In our computation, we set a maximal tolerance $\mathcal{E}$ of the fluctuation of $\mathrm{Im}(\hat{S}(\vec{z}))$. The tolerance determines the size of $\hat{V}_\sigma$, such that at any point $p\in\hat{V}_\sigma$, $|\mathrm{Im}(\hat{S}(p))-\mathrm{Im}(\hat{S}(p_\sigma))|\leq \mathcal{E}$. Similarly, $\arg(\det(J))$ does not have strong fluctuation in our case of the spinfoam expectation values. When the integrands are weakly oscillatory functions, both $\langle \e^{\ii\theta_{res}} \hat{f} \rangle_{eff}$ and $\langle \e^{\ii\theta_{res}} \rangle_{eff}$ can be  accurately computed by the MCMC method.

\subsection{Differential Evolution Adaptive Metropolis (DREAM) Algorithm}

By now, we have converted the problem of computing $\langle f \rangle$ to the problem of how to sample on the tangent space $\hat{V}_{\sigma'}$ with a Boltzmann factor $\e^{- {S}_{eff}}$ to compute $\langle\e^{\ii \theta_{res}}\rangle_{eff}$ and $\langle\hat{f}\e^{\ii \theta_{res}}\rangle_{eff}$. The latter problem can be numerically solved by Markov Chain Monte Carlo (MCMC) method \cite{brooks_meng_jones_galman_2011}. 

The MCMC is a class of algorithms designed for sampling from a posterior probability distribution $\e^{- {S}_{eff}}$. Each Markov chain can be regarded as generated by a random `walker' moving in the integration domain. In general, in each step of the `walker', MCMC methods use the accept/reject scheme to adjust the transition distribution that dictates the orientation and the length of the step, such that the points sampled by the `walker' converge to a desired {posterior} distribution after a `long march'. 

Since in our computation of the integrals \eqref{eq:core4} - \eqref{eq:core7}, $\e^{- {S}_{eff}}$ is high-dimensional and complicated, we choose a MCMC method---the Differential Evolution Adaptive Metropolis (DREAM)\cite{DREAM, VRUGT2016273}. The DREAM algorithm has not been combined with the Lefschetz thimble method in the literature.

The DREAM algorithm runs multiple Markov chains in parallel. This algorithm is able to sample in different regions of the integration domain simultaneously. In case that ${S}_{eff}$ has some local minima other than the critical point (global minimum), unlike the one-chain scheme, this multi-chain scheme prevents a sampling procedure from being trapped in the neighborhood of any local minimum \cite{DREAM}. Besides, this multi-chain scheme is more adaptable to the architecture of high performance computers designed for multitasking.

The performance of a MCMC method also depends on the quality of the candidates of each update provided by the method. On the one hand, at each update, if a candidate is too far from the current location of the chain, the candidate is rejected and wastes the computation resource. On the other hand, if a candidate is too close to the current location of the chain, although it is highly probable to be accepted, the Markov chain may take numerous updates to explore the whole integration domain. In order to balance the progress of the chain in each step and a reasonable acceptance rate, the DREAM algorithm possesses the following features 
\begin{itemize}
  \item At each update, the DREAM uses a genetic algorithm to provide a candidate for each chain, based on the current location of other chains.
  \item Similar to the Gibbs sampling, the DREAM does not update all the components in one update. Instead, it implements a randomized subspace sampling strategy. For each update of each chain, the algorithm does not update all the components of the sample. There is a probability, called the crossover ratio $CR$, to decide whether a component needs to be updated or not.
  \item The algorithm suggests to do test runs, called burn-in runs, before the formal sampling. In the burn-in runs, the crossover ratio $CR$ and the parameters used in the genetic algorithm are adapted. Furthermore, in the burn-in runs, the unwanted chains can be removed by using inter-quartile range method.
\end{itemize}

Explicitly, after a proper burn-in run, we can compute the $\langle f \rangle$ in the following steps. 

\begin{enumerate}

  \item Compute the Hessian of the action $\hat{S}(\vec{z})$ at the critical point $p_{\sigma}$. Use \eqref{eq:HES2} to compute the basis vectors $\{ \hat{\omega}_i \}$ of the thimble's tangent space $\hat{V}_{\sigma}$. As claimed before, the action $\hat{S}(\vec{z})$ is defined in $\mathbb{C}^n$, and $\hat{V}_{\sigma}$ is real $n$-dimensional. 
   
  \item Choose $M$ points in $\mathbb{R}^n$ close to the critical point \footnote{As suggested in \cite{DREAM}, $M$ should be greater than or equal to $n$}, and denote them as $\mathbf{x}_s^{(0)},\ (s=1,\cdots, M)$, where the index $(0)$ indicates the initial step. Then, ${\mathbf{x}_s^{(0)}}\cdot\hat{\omega},\ (s=1,\cdots, M)$ are the initial points of the $M$ Markov chains on $\hat{V}_{\sigma}$ (some more details of our choice of the initial points are shown in {Section} \ref{sec:opt}).
  
  \item Generate $CR$ from a given multi-nomial distribution constructed in a burn-in run. Construct a $n$-element series $\{u^i\}$, where each $u^i$ is drawn from a uniform distribution $U(0,1)$. Then, construct a $n$-dimensional vector $\mathbf{v}$ whose components are given by
  \[
  v^i=
    \begin{cases}
      0&\text{If $u^i > CR$};\\
      1& \text{otherwise.}\\
    \end{cases}
  \] 
  \item Create $x_s^{cand}$ of each chain by  
  \be\label{eq:dreamcandidate}
  \mathbf{x}_s^{cand}=\mathbf{x}_s^{(t-1)}+\mathbf{v}\cdot(\mathbf{1}_n+\mathbf{e})\beta(\delta,d')(\sum_{j=1}^{\delta}\mathbf{x}_{R_1(j)}^{(t-1)}-\sum_{d=1}^{\delta}\mathbf{x}_{R_2(d)}^{(t-1)})+\mathbf{\epsilon},\ s=1,\cdots M, 
  \ee
  where $d'$ is the total number of non-zero components in $\mathbf{v}$, $\delta$ is the number of the pairs used to generate the candidates, and $R_1(j),\ R_2(d)\in\{1,\cdots,M \}$, with $R_1(j)\neq R_2(d)\neq s$ for $j=1,\cdots \delta$ and $d=1,\cdots, \delta$.  The values of $\mathbf{\epsilon}$ and $\mathbf{e}$ are drawn from the Gaussian distribution $N(0,b^*)$ and uniform distribution Uniform$(-b,b)$ with $|b|<1$ respectively. The parameter $b^*$ is chosen to be very small compared with the width of the target posterior distribution. The scaling factor is $\beta(\delta,d')$. At every fifth generation, we set $\beta=1$.  Denote the candidate update for the $s$-th chain, i.e., $\mathbf{x}_s^{cand}\cdot\hat{\omega}$, as $\hat{\mathbf{x}}_s^{cand}$.
  \item Compute $\tilde{\mathbf{x}}_s^{cand}=\mathcal{C}_T(\hat{\mathbf{x}}_s^{cand})$ as the candidate sample on the thimble, and compute the corresponding Jacobian by solving \eqref{eq:jacobian}. Then, use the Jacobian and $\tilde{\mathbf{x}}_s^{cand}$ to compute $S_{eff}(\hat{\mathbf{x}}_s^{cand})$.
  \item For each chain, construct an acceptance rate as 
  \[
    r_s^{(t)}=\min\left( 1, \frac{S_{eff}(\hat{\mathbf{x}}_s^{cand})}{S_{eff}(\hat{\mathbf{x}}_s^{(t-1)})} \right).
  \]
  Nevertheless, The following two exceptions where we let the rate becomes $0$ exist: 
  \begin{itemize}
    \item when $|{\rm Im}(\hat{S}(\hat{\mathbf{x}}_s^{cand}))-{\rm Im}(\hat{S}(\hat{\mathbf{x}}_s^{0}))|>\mathcal{E}$, where $\mathcal{E}$ is the preset tolerance of the fluctuation of ${\rm Im}(\hat{S})$;
    \item when numerically $\mathcal{C}_T(\hat{\mathbf{x}}_s^{cand})$ cannot be accurately computed. 
  \end{itemize}
  The second exception is due to that the SA equations become unstable when it flows to the region far from the critical point; however, such far points contribute very little to the final result due to the exponential suppression $e^{\mathrm{Re}(\hat{S})}$. Hence, we ignore the contributions of such far points. 
     
  \item Create a number $a\in[0,1]$ from uniform distribution, and then use it to decide whether a candidate can be accepted or not:
  \[
    x_s^{(t)}=
  \begin{cases}
    x_s^{cand}&\text{If $a<r_s^{(t)}$},\\
    x_s^{(t-1)}&otherwise.\\
  \end{cases}  
  \]
  \item Return to step 3 until a sufficiently large number of samples are collected.
  \item Use the collected samples to compute 
  \[
  \langle f \rangle = \frac{\langle \e^{\ii\theta_{res}} \hat{f} \rangle_{eff}}{\langle \e^{\ii\theta_{res}} \rangle_{eff}}.
  \]
When we have a large number of samples, $\langle \mathcal{O} \rangle_{eff}$ is equivalent to arithmetic mean among the samples
\be
\langle \mathcal{O} \rangle_{eff}=\sum_{\rm samples}\mathcal{O}({\rm sample}).
\ee
since the above procedure has produced the desired probability distribution $\e^{- {S}_{eff}}$ for the samples. 
  
\end{enumerate}


We provide a more detailed discussion of the MCMC methods in \textbf{Appendix} \ref{apd:MCMC}. In order to perform better, several optimizations exist, such as choosing a better the flow-time $T$, effectively solving the SA equations to get $\mathcal{C}_T$, and properly tuning the multi-nomial distribution of $CR$ and the scale factor $\beta$ during the burn-in runs. We provide the details of these optimizations in Section \ref{sec:opt}.

\section{Spinfoam on Lefschetz Thimble}
In the above section, we have described the general algorithm of integrals on Lefschetz thimbles. In this section, we apply the Lefschetz thimble to the spinfoam model. 

     
We need to first complexify the spinfoam variables $j_{ab}, g_a, z_{ab}$ and analytically continue the integrands in \eqref{eq:core4} - \eqref{eq:core7}. The analytic continuation makes $g^\dagger_a$ and $\bra{z_{ab}}$ independent of $g_a$ and $\ket{z_{ab}}$. Equivalently, using the parametrizations \eqref{eq:par1} and \eqref{eq:par2}, we complexify $\vec{x}$ and $\vec{y}$ and analytically continue the integrands to holomorphic functions of $\vec{x}$ and $\vec{y}$. The spin variables $j_{ab}$ are also complexified, and the integrands are holomorphic in $j_{ab}$. The real scaling parameter $\lambda$ is kept real. The analytical continuation render the integrands and in particular, the action denoted by $\tilde{S}_{tot}$, holomorphic functions of $54$ complex variables. The thimble is a real $54$-dimensional sub-manifold in the space of complexified spinfoam variables. Note that the analytic continuation of the spinfoam integrand has been discussed in \cite{LowE1} (see also \cite{toappear}). 
     
After analytical continuation, $\tilde{S}_{tot}$ may have more critical points than $S_{tot}$ does. Complex critical points may exist in addition to the real critical points discussed above. The complex critical points are away from the real integration domain, where the spinfoam integrals \eqref{eq:core4} - \eqref{eq:core7} are defined. These spinfoam integrals admit decompositions as in \eqref{eq:decomT}, where $\sigma$'s may contain both real and complex critical points. A complex critical point $p_{\tilde{\sigma}}$ contributes to the integrals if $n_{\tilde \sigma}\neq 0$, i.e. there exist SA paths approaching $p_{\tilde{\sigma}}$ from the space of real variables. Thus, $n_{\tilde \sigma}\neq 0$ implies that $\mathrm{Re}(\tilde{S}_{tot}(p_{\tilde{\sigma}}))>0$ because on the space of real variables, $\mathrm{Re}(\tilde{S}_{tot})=\mathrm{Re}({S}_{tot})\geq 0$. Strictly positive $\mathrm{Re}(\tilde{S}_{tot}(p_{\tilde{\sigma}}))$ implies that when the spinfoam integrals are decomposed as in Eq. \eqref{eq:decomT}, $\mathcal{J}_{\tilde{\sigma}}$ contributes exponentially small at large $\lambda$.

At large $\lambda$, the single geometrical critical point $p_{geo}$ dominates the spinfoam integrals \eqref{eq:core4} - \eqref{eq:core7}, as discussed in Section \ref{subsec:critical}. Hence, the single Lefschetz thimble $\mathcal{J}_{geo}$ associated to $p_{geo}$ dominates the decomposition \eqref{eq:decomT} of the spinfoam integrals. Therefore, \eqref{eq:regThimble} is applicable to the expectation values in (\ref{expvalues}) at large $\lambda$. As a result, we pass from \eqref{eq:core4} - \eqref{eq:core7} to integrals on $\mathcal{J}_{geo}$ 
\begin{align}
  \langle W|\Psi_0\rangle&\simeq  \int_{\mathcal{J}_{geo}} \mathrm{d} j \mathrm{d} \phi \,\hat{U}  \,\e^{-\lambda \hat{S}_{tot} },\label{eq:core4f}\\
  \bra{W}E^a_n\cdot E^b_n E^c_m \cdot E^d_m |\Psi_0\rangle&\simeq \int_{\mathcal{J}_{geo}} \mathrm{d} j \mathrm{d} \phi\,  \hat{U} \, \e^{-\lambda \hat{S}_{tot} } (\hat{A}_{an} \cdot \hat{A}_{bn}) (\hat{A}_{cm} \cdot \hat{A}_{dm}),\label{eq:core5f}\\
  \bra{W}E^a_n\cdot E^b_n |\Psi_0\rangle&\simeq \int_{\mathcal{J}_{geo}} \mathrm{d} j \mathrm{d} \phi\, \hat{U}\, \e^{-\lambda \hat{S}_{tot} } (\hat{A}_{an}  \cdot \hat{A}_{bn} ),\label{eq:core6f}\\
  \bra{W} E^c_m \cdot E^d_m |\Psi_0\rangle& \simeq \int_{\mathcal{J}_{geo}} \mathrm{d} j \mathrm{d} \phi\,  \hat{U} \, \e^{-\lambda \hat{S}_{tot}}  (\hat{A}_{cm}  \cdot \hat{A}_{dm} )\label{eq:core7f}, 
\end{align}
where all $\hat{U}, \hat{S}_{tot}, \hat{A}_{an}$ are holomorphic functions of complexified spinfoam variables. The expectation values in (\ref{expvalues}) reduce to the desired forms as $\langle f \rangle$ in \eqref{eq:regThimble} and can be computed by MCMC methods.

Equations \eqref{eq:core4f} - \eqref{eq:core7f} capture the contributions of the dominant critical point $p_{geo}$ to (\ref{eq:core4}) - \eqref{eq:core7} and include all orders of perturbative $1/\lambda$ corrections. Namely, when we expand (\ref{eq:core4}) - \eqref{eq:core7} as $1/\lambda$ power series at the critical point $p_{geo}$, under the stationary phase approximation, the power series are the same as expanding  (\ref{eq:core4f}) - \eqref{eq:core7f} in $1/\lambda$ (see e.g. \cite{Cristoforetti:2012su} for a general argument). Nevertheless, the approximation leading to (\ref{eq:core4f}) - \eqref{eq:core7f} from (\ref{eq:core40}) - \eqref{eq:core70} neglect the contributions that are exponentially suppressed at large $\lambda$. These contributions are (1) integrals with $k_{ab}\neq 0$ in (\ref{eq:core40}) - \eqref{eq:core70}, (2) extending some integrals to infinite such as $\int \dd j_{ab} $ and $\int\dd \phi $ on the cover space, and (3) the complex critical points and corresponding Lefschetz thimbles. 

What we have shown so far is that each quantity in (\ref{eq:core0}), (\ref{eq:core1}) - \eqref{eq:core2} and the spinfoam propagator can be expressed as the power series $\sum_s a_s\lt(\frac{1}{\lambda}\rt)^s$ plus contributions exponentially suppressed (or namely suppressed faster than $O(1/\lambda^N)$ for any integer $N$) at large $\lambda$. Eqs. \eqref{eq:core4f} - \eqref{eq:core7f} capture the power series while neglecting the exponentially suppressed contributions. The power series contain all the perturbative quantum corrections. The exponentially suppressed contributions may be called non-perturbative corrections, as they contain the sub-dominant thimbles associated with the complex critical points generated by the analytical continuation. In this language, Eqs. \eqref{eq:core4f} - \eqref{eq:core7f} capture all perturbative quantum corrections in (\ref{eq:core0}), (\ref{eq:core1}) - \eqref{eq:core2} while neglecting non-perturbative corrections.

It is known that in the traditional stationary phase expansion, the computational complexity grows exponentially when computing $a_s$---the coefficient of $O(1/\lambda^s)$ correction---with larger $s$, so it is very difficult to sum the power series $\sum_s a_s\lt(\frac{1}{\lambda}\rt)^s$ in the traditional approach. In this sense, our method with the Lefschetz thimble is a powerful way to compute the spinfoam propagator containing perturbative quantum corrections to all orders.

Besides, similar to the idea in \cite{Witten:2010zr,Cristoforetti:2012su}, we can consider the integral (\ref{eq:core4f}) with the Lefschetz thimble as a new definition of the spinfoam model. When generalizing to arbitrary simplicial complex $\mathcal{K}$, we define the spinfoam model on Lefschetz thimble by
\be
Z_{\mathcal{J}}=\int_{\mathcal{J}}\dd j\dd\phi\, \hat{U}_{\mathcal{K}}\,e^{-\lambda \hat{S}_{\mathcal{K}}},\label{ZJ}
\ee
where $\hat{S}_{\mathcal{K}}$ is the analytic continuation of the spinfoam action on $\mathcal{K}$ \cite{LowE1}. Here, $\int\dd j$ integrates all internal spins, and $\mathcal{J}$ is the Lefschetz thimble associated with a single critical point. Applying the Lefschetz thimble to spinfoam model has been proposed earlier in the context of coupling to cosmological constant \cite{Haggard:2015yda,Haggard:2015kew,hanSUSY}. Eq. (\ref{ZJ}) has the advantage to focus on the contributions from a single critical point and excludes other critical points. In particular, when $\mathcal{J}$ is the thimble of the critical point corresponding to the Lorentzian Regge geometry, it excludes contributions from vector geometries and the geometries with flipping orientations (see e.g. \cite{HZ,hanPI} for the classification of critical points). In addition, Eq. (\ref{ZJ}) is a better formulation from the computational point of view, as the main point of this paper. 

Given the critical point, the spinfoam model on Lefschetz thimble has the same perturbative $1/\lambda$ expansion as the usual definition of the spinfoam amplitude and in particular has the same semi-classical limit as the usual spinfoam amplitude, as shown in the numerical results in Section \ref{Numerical Results}. The small $\lambda$ behavior of (\ref{ZJ}) is different the usual spinfoam amplitude because non-perturbative corrections are not negligible at small $\lambda$.

\section{Optimizations}\label{sec:opt}

In this section, we provide some technical details of the optimizations used in our computation. 

\subsection{Optimizations of Solving Steepest Ascent (SA) Equations}
One crucial step of our algorithm is to solve the SA equation \eqref{eq:SA} for a given initial condition. For spinfoam model, the action is written as $\lambda\tilde{S}_{tot}$ after analytic continuion ($\lambda\tilde{S}_{tot}$ plays the role of $\hat{S}(\vec{z})$). The right hand side of the SA equation is proportional to the scaling parameter $\lambda$. The idea of the ODE numerical solvers is to use a difference equation to approach the given differential equation. In our case, the difference equation is given by
\[
  \frac{\dd z^a }{ \dd t}=\lambda\frac{\partial\overline{\tilde{S}_{tot}}}{\partial\overline{z^a}}
  \sim
  \Delta  z^a=\lambda \frac{\partial\overline{\tilde{S}_{tot}}}{\partial\overline{z^a}}\Delta t,
\]
where the $\Delta  z^a$ in the left hand side is supposed to be small to bound the numerical error at each update. For a fixed error tolerance in each time step, the time step $\Delta t$ has to be small at large $\lambda$, or at large $|\partial\tilde{S}_{tot}/\partial z^a|$. This fact has two indications. First, at large $\lambda$, in order to keep the accuracy, the total evolution can not be too long. Second, at large $|\partial\tilde{S}_{tot}/\partial z^a|$, the numerical solver may be inaccurate. 

In view of the first implication, we would let the total evolution time $T$ be an element of $\{ \tau /\lambda| \tau\in[0.1,1]\}$. In our computation, we have $\lambda\in[10^2, 10^7]$ and set the tolerance $\mathcal{E}=0.1/\lambda$ of the fluctuation of $\mathrm{Im}(\hat{S}_{tot})$. The tolerance $\mathcal{E}$ determines the shape and size of $\hat{V}_{\sigma}$. The value of $T$ determines the size the $\tilde{\mathcal{J}}_\sigma$ in which the MCMC is actually carried out. Thanks to the fast decaying $\e^{-\lambda \mathrm{Re}(\hat{S}_{tot})}$ when the Markov chains are moving far from the critical point $p_{\sigma}$, we find that a relatively small $T$ is sufficient to generate a $\tilde{\mathcal{J}}_\sigma$ large enough, where the Markov chain can sample all dominantly contributing points. The points outside $\tilde{\mathcal{J}}_\sigma$ contribute exponentially small and are neglected.

The second implication usually happens when the initial point $x$ is not close to the critical point. In this case, $\mathcal{C}_T(x)$ is very far away from the critical point and $|\partial\tilde{S}_{tot}/\partial z^a|$ at $\mathcal{C}_T(x)$ will be large. The real part ${\rm Re}{(\tilde{S}}_{tot})$ is large at $\mathcal{C}_T(x)$. These points contribute exponentially small to the final integral and are negligible. In our work, we use the embedded Runge-Kutta-Fehlberg Method (RKF45) to be the numerical solver of the steepest ascent equation \eqref{eq:SA}\footnote{In principle, other embedded Runge-Kutta should also work.}. The solver can automatically determine the step size based on the given tolerance of the total error. In the accept/reject step of the DREAM algorithm, we can directly reject a candidate $x^{cand}$ if one of the following two events occurs in solving $\mathcal{C}_T(x^{cand})$
\begin{itemize}
  \item The adaptive ODE step size $\Delta t$ is smaller than a threshold.
  \item $|{\rm Im}{(\tilde{S}_{tot}(\mathcal{C}_T(x^{cand})))}-{\rm Im}{(\tilde{S}_{tot}(x^{cand}))}|$ exceeds the tolerance $\mathcal{E}$. 
\end{itemize}
To be concrete, we set the tolerance of the total error for the ODE as $10^{-20}$, set the threshold for the smallest ODE step size as $10^{-6}$, and the maximum error tolerance on the imaginary part of the action as $0.1$.

  

\subsection{Optimizations of the DREAM}
To make our sampling procedure in the DREAM algorithm more efficient, we can optimize the choice initial points, the evolution time $T$, and burn-in runs.

\subsubsection{Optimizing the choice of the initial points}

In our work, we use $108$ parallel chains to run the DREAM algorithm, so we need $108$ initial points at the beginning. Although in principle, the initial points can be randomly chosen, a good choice may possibly reduce the number of burn-in runs to save time and computational resources. Our choice of the initial points is given by the following scheme:
\begin{itemize}
  \item Compute the basis vectors $\{ \hat{\omega}_i \}$ of the thimble's tangent space,
  \item For each $\hat{\omega}_i$, use enumeration method to find a number $\eta_i$ so that ${S}_{eff}(\eta_i\hat{\omega}_i)$ is valued in between $1$ and $0.1$,
  \item Use $\{ \pm\eta_i\hat{\omega}_i;\ i=1,\cdots,54 \}$ as the initial points.
\end{itemize} 
Since we treat $\e^{-{S}_{eff}}$ as a Boltzmann factor, we can consider ${S}_{eff}$ as the energy. By this scheme, our choice covers more directions and synchronizes the energy of the initial points.

\subsubsection{Optimizing the flow time $T$ }

\begin{figure}[ht]
  \centering
  \subfigure[Absolute values ($\lambda=100$)]{
  \begin{minipage}[t]{0.45\linewidth}
  \centering
  \includegraphics[width=3.2in]{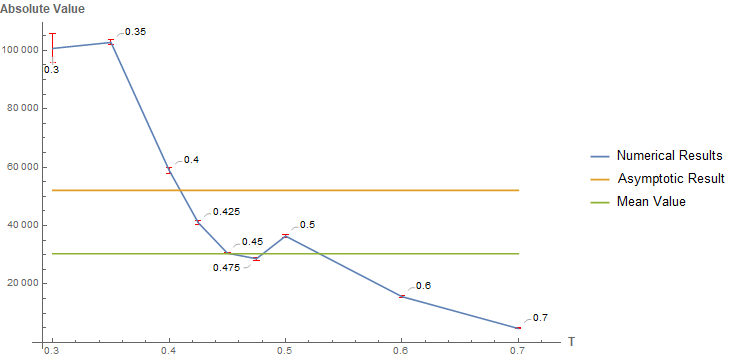}
  \end{minipage}%
  }%
  \subfigure[Arguments ($\lambda=100$)]{
  \begin{minipage}[t]{0.45\linewidth}
  \centering
  \includegraphics[width=3.2in]{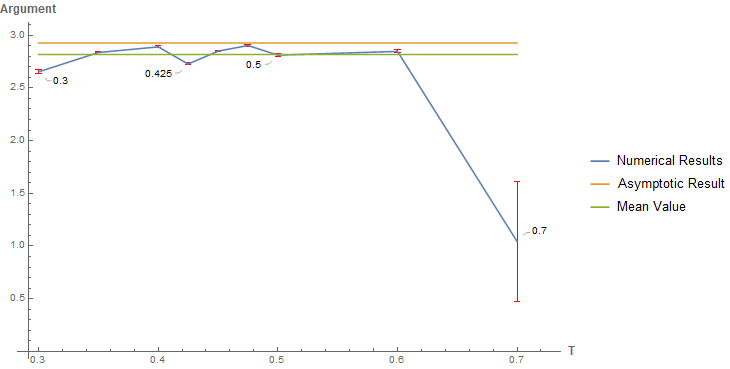}
  \end{minipage}%
  }%
  \centering
\caption{The absolute values and arguments of $G^{2315}_{14}\ (\lambda=100)$ computed by the thimble algorithm with different $T$.}\label{fig:T1}
\end{figure}

\begin{figure}[htbp]
  \centering
  \subfigure[Absolute values ($\lambda=10^4$)]{
  \begin{minipage}[t]{0.5\linewidth}
  \centering
  \includegraphics[width=3.5in]{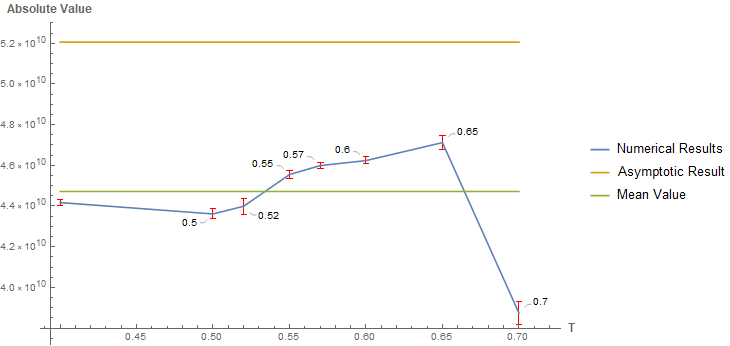}
  \end{minipage}%
  }%
  \subfigure[Arguments ($\lambda=10^4$)]{
  \begin{minipage}[t]{0.5\linewidth}
  \centering
  \includegraphics[width=3.5in]{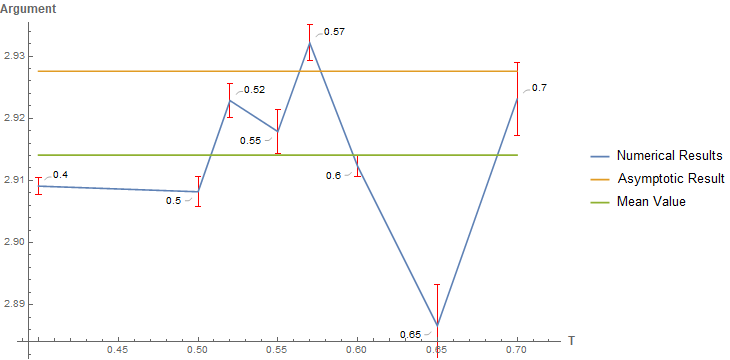}
  \end{minipage}%
  }%
  \centering
\caption{The absolute values and arguments of $G^{2315}_{14}\ (\lambda=10^4)$ computed by the thimble algorithm with different $T$.}\label{fig:T2}
\end{figure}

The flow time $T$ is an undetermined parameter of the algorithm. The setting of $T$ cannot be too large or too small. If $T$ is too small, the fluctuation of ${\rm Im}(\tilde{S}_{tot})$ may be too large to keep the result accurate. If $T$ is too large, only a very small portion of $\hat{V}_\sigma$ contributes to the integrals because $\e^{-{S}_{eff}}$ decays faster at larger $T$. 
Then the shape of the target distribution $\e^{-{S}_{eff}}$ is too sharp when $T$ is too large. In principle, in contrast to the `moderate' distributions, this sharp distribution can only be approximated by a longer Markov chain. In other words, too large $T$ make the algorithm less efficient. In practice, we suggest the following scheme to deal with the choice of $T$:

\begin{itemize}
  \item Run the algorithm with several trials with different time $T$.
  \item Update the Markov chains in each $T$-trial similar number of times.
  \item Sort the computational results of the propagator (or at least one component of the propagator) from the $T$-trials in decreasing order of $T$.
  \item The results from many $T$-trials are close to one another. As such, increasing $T$ only adds negligible contribution to the integral. We take the mean value of these results as the final result. 
\end{itemize}
For example, we show our results of the component $G^{2315}_{14}$ of the spinfoam propagator, in the case of $\lambda=100$ and $\lambda=10^5$ in FIG. \ref{fig:T1} and FIG. \ref{fig:T2} respectively. 

In the case where $\lambda=100$, the set of flow times $T=\{0.3,0.35,0.4,0.425,0.35,0.475,0.5,0.6,0.7\}$ is used for the computations. In each $T$-trial, the value of $G^{2315}_{14}$ is computed based on roughly $10^7$ samples. The results on the blue line in FIG. \ref{fig:T1} indicate that, given the number of samples, the results from the $T$-trials with $T=\{0.425,0.35,0.475,0.5,0.6\}$ are close to one another, so their mean value $(28373\pm610)\e^{\ii(2.83\pm0.006)}$ (shown as the green line in FIG. \ref{fig:T1}) is taken as the numerical result of $G^{2315}_{14}$. This result has $(37.90\pm0.6)\%$ percentage difference comparing with the result from the asymptotic expansion up to $O(1/\lambda)$ (shown as the yellow line in FIG. \ref{fig:T1}). 

Similarly, in the case where $\lambda=10^4$, we choose the set of flow times $T=\{0.4, 0.5, 0.52, 0.55, 0.57, 0.6, 0.65, 0.7\}$. FIG. \ref{fig:T2} depicts the results based on over $10^6$ samples\footnote{Compared with the case of $\lambda=100$, $10^6$ is already a sufficiently large number of samples to make the results at $\lambda=10^4$ converge because the contributed region in $\hat{V}$ at $\lambda=10^4$ is smaller and thus easier to be simpled than that at $\lambda=100$.}. Except the results from the $T$-trials with $T=\{0.4, 0.7\}$, the results are close to one another and their mean value is $((4.548\pm0.015)\times10^{10})\e^{\ii (2.913\pm0.0017)}$ (the green line in FIG. \ref{fig:T2}). This value differs from that due to the asymptotic expansion (the yellow line in FIG. \ref{fig:T2}) by $(13.22\pm0.28)\%$ of the latter.

\subsubsection{Optimizing the burn-in stage}
As mentioned before, the multi-nomial distribution determining the crossover ratio $CR$ and the scale factor $\beta$ of the DREAM can be tuned during the burn-in runs. Following the Ref.\cite{DREAM}, one can adopt \textbf{Algorithm} \ref{alg:burn-in} in the burn-in runs to tune the $\text{multi-nomial distribution} (.\ ;p_1,\cdots,p_m)$.  
\begin{figure}[h]
  \begin{algorithm}[H]
    \caption{DREAM burn-in}\label{alg:burn-in}
    \begin{algorithmic}[1]
      \STATE initial $t\gets 1$, $L_m\gets 0$, $p_m=1/n_{cr}, m=1,\cdots,n_{cr}$
      \WHILE {burn-in steps $t<K$}
      \FOR {chains $i=1,\cdots, M $}
      \STATE $m\sim\text{multinomial}(.\ ;p_1,\cdots,p_m)$
      \STATE $CR\gets m/n_{CR}$ and $L_m=L_m+1$
      \STATE Create a candidate 
      \STATE Accept/Reject the candidate
      \STATE $\Delta_m\gets\Delta_m+\sum_{j=1}^d((x_i^{(t)})^j-(x_i^{(t-1)})^j)^2/r_j^2$, where $r$ denotes the standard deviation current locations of the chains.
      \ENDFOR
      \STATE $p_m \gets tN\cdot(\Delta_m/L_m)/\sum_{j=1}^{n_{CR}}\Delta_j$
      \STATE $t\gets t+1$
      \ENDWHILE
      \end{algorithmic}
  \end{algorithm}
  \end{figure}

For the scale factor, as mentioned in Ref.\cite{DREAM}, if the target distribution is Gaussian, the optimal $\beta=2.4/\sqrt{2d'\delta}$ yields the acceptance rate equal to $0.44$ for $d'=1$, around $0.2$ for larger $d'$. But based on our test, this choice is not suitable for our case. In order to optimize the performance of the algorithm, we tune $\beta$ based on the acceptance rate during the burn-in runs. At the beginning of the burn-in stage, we set $\beta=2.4C/\sqrt{2d'\delta}$ with $C=1$. In each update during the burn-in, we count the number of accepted candidates as $\alpha$, then we compute the acceptance rate in this update as $\alpha/108$. If the acceptance rate is greater than $0.4$, we multiply $C$ by $1.2$, whereas if the acceptance rate is greater than $0.1$, we multiply $C$ by $0.5$\footnote{The threshold and the multiple rate here is chosen by hand. One may need to adjust them for other computing tasks.}. After a long burn-in stage, we expect that the $\beta$ is tuned such that the acceptance rate is round $0.3$, which is a optimal value for high-dimensional problem.

\section{The Large Spin Limit of Spinfoam Propagator}

In this section, we depart from Lefschetz thimbles and discuss the standard stationary phase analysis in the large-$\lambda$ limit of the spinfoam propagator. We are going to compare between the large-$\lambda$ limit the results from MCMC on Lefschetz thimble in Section \ref{Numerical Results}.

As in Refs.\cite{propagator,propagator1,propagator2,propagator3}, the spinfoam propagator can be computed by the asymptotic expansion following the stationary phase analysis in Ref.\cite{stationaryphase}:
\be\label{eq:HOM}
\left|\int_{K} u(x) e^{i \lambda f(x)} d x-e^{i \lambda f\left(x_{0}\right)}\left[\operatorname{det}\left(\frac{\lambda f^{\prime \prime}\left(x_{0}\right)}{2 \pi i}\right)\right]^{-\frac{1}{2} } \sum_{s=0}^{k-1}\left(\frac{1}{\lambda}\right)^{s} L_{s} u\left(x_{0}\right)\right| \leq C\left(\frac{1}{\lambda}\right)^{k} \sum_{|\alpha| \leq 2 k} \sup \left|D^{\alpha} u\right|,
\ee
with
\[
g_{x_{0}}(x)=f(x)-f\left(x_{0}\right)-\frac{1}{2} H^{a b}\left(x_{0}\right)\left(x-x_{0}\right)_{a}\left(x-x_{0}\right)_{b},
\]
and
\[
L_{s} u\left(x_{0}\right)=i^{-s} \sum_{l-m=s} \sum_{2 l \geq 3 m} \frac{(-1)^{l} 2^{-l}}{l ! m !}\left[\sum_{a, b=1}^{n} H_{a b}^{-1}\left(x_{0}\right) \frac{\partial^{2}}{\partial x_{a} \partial x_{b}}\right]^{l}\left(g_{x_{0}}^{m} u\right)\left(x_{0}\right),
\]
where $H(x)=f''(x)$ is the Hessian matrix.
Using the parametrization mentioned in {Section} \ref{sec:SFP}, we use \textit{Mathematica}\texttrademark to compute the Hessian of the spinfoam action and derivatives of $A^i_{ab}$ and $U$ defined in {Section} \ref{subsec:SFP}.  Then following (\ref{eq:HOM}), we compute the $1/\lambda$ expansion of $\left\langle E^a_n\cdot E^b_n E^c_m\cdot E^d_m \right\rangle$, $\left\langle E^a_n\cdot E^b_n\right\rangle$, and the spinfoam propagator $G^{abcd}_{mn}$. The code of this computation is shared in \cite{hzcgit1}. The results are used as the reference data in comparison with the Lefschetz thimble Monte-Carlo computations. 

For example, if we keep the expansion \eqref{eq:HOM} to the first order of $1/\lambda$ (keeping the terms of $s=0,1$), the components $\langle E^2_1\cdot E^3_1 E^1_4 \cdot E^5_4\rangle$, $\langle E^1_4 \cdot E^5_4\rangle$, $\langle E^2_1\cdot E^3_1\rangle$ and $G^{2315}_{14}$ are given by 
\be\label{eq:DEE1}
\begin{split}
\frac{\langle E^2_1\cdot E^3_1 E^1_4 \cdot E^5_4\rangle}{\lambda^4} = 0.006944+\frac{0.03659-0.009716\ii}{\lambda} +O\lt(\frac{1}{\lambda^2}\rt),
\end{split}
\ee
\be\label{eq:DEE2}
\frac{\langle E^2_1\cdot E^3_1\rangle }{\lambda^2}= -0.08333+\frac{2.292-0.5092\ii}{\lambda}+O\lt(\frac{1}{\lambda^2}\rt)
\ee
\be\label{eq:DEE3}
\frac{\langle E^1_4 \cdot E^5_4\rangle}{\lambda^2} = -0.08333+\frac{1.242-0.2599\ii}{\lambda}+O\lt(\frac{1}{\lambda^2}\rt),
\ee
\be\label{eq:DEE4}
G^{2315}_{14}\sim -(0.05087-0.01106\ii)\lambda^3+O(\lambda^2)
\ee
We choose the Barbero-Immirzi parameter as $\gamma=-0.1$ in the above numerics.

The $\lambda^3$ behavior of the spinfoam propagator can be seen analytically, by the known leading order formula \cite{propagator,propagator1,propagator2,propagator3,Han:2017isy}: 
\be\label{eq:lead}
G^{a b c d}_{mn}\sim \lambda^{-1} H^{-1}_{\alpha\beta} \partial_\alpha(A_{an}\cdot A_{bn}) \partial_\beta(A_{bm}\cdot A_{dm}),
\ee
where $H$ is the Hessian of $S_{tot}$ at the critical point $p_{geo}$, and the indices $\alpha,\, \beta$ correspond to the variables $j$ and $\phi$ defined in Section \ref{sec:SFP}. By definition \eqref{eq:ADE}, $A_{an}$ is proportional to $\lambda$, so \eqref{eq:lead} is at the order of $\lambda^3$. This result turns out to be consistent with our numerical result from the Lefschetz thimble Monte-Carlo in the large-$\lambda$ limit.

     
\section{Numerical Results}\label{Numerical Results}

Recall \eqref{eq:prop1} that the spinfoam propagator is obtained by computing the expectation values $\left\langle E^a_n\cdot E^b_n E^c_m\cdot E^d_m \right\rangle$ and $\left\langle E^a_n\cdot E^b_n\right\rangle$. We choose the {Barbero-Immirzi parameter} $\gamma=-0.1$ in this computation and compute these expectation values and the propagator in the situations where $\lambda=50$, $100$, $1000$, $10000$, $50000$, $100000$, $500000$, $1000000$, $5000000$, $10000000$ and $50000000$. In the first and second parts of this section, we show the numerical results respectively of the expectation values and of the propagator. 

\subsection{Expectation Values}

\begin{figure}[htbp]
  \centering
  \subfigure[Absolute value versus $\lambda$]{
  \begin{minipage}[t]{0.5\linewidth}
  \centering
  \includegraphics[width=3.5in]{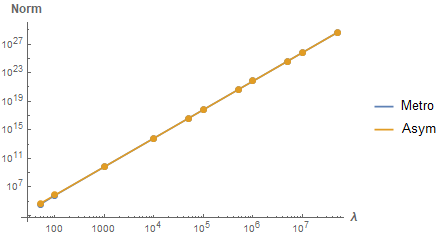}
  \end{minipage}%
  }%
  \subfigure[Argument versus $\lambda$]{
  \begin{minipage}[t]{0.5\linewidth}
  \centering
  \includegraphics[width=3.5in]{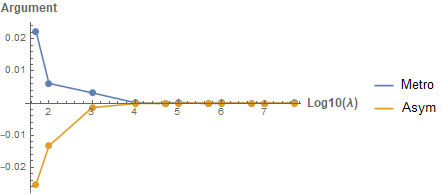}
  \end{minipage}%
  }%
  \centering
\caption{The absolute values and the arguments of $\langle E^2_1\cdot E^3_1 E^1_4 \cdot E^5_4\rangle$. The results of the asymptotics are shown in the yellow lines, and the numerical results are indicated by blue line. }\label{fig:EE}
\end{figure}

\begin{figure}[htbp]
  \centering
  \subfigure[Absolute value versus $\lambda$]{
  \begin{minipage}[t]{0.5\linewidth}
  \centering
  \includegraphics[width=3.5in]{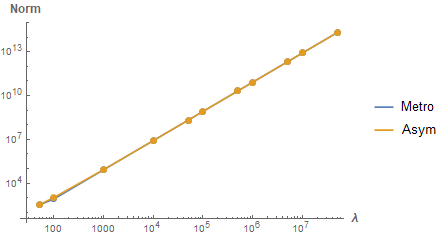}
  \end{minipage}%
  }%
  \subfigure[Argument versus $\lambda$]{
  \begin{minipage}[t]{0.5\linewidth}
  \centering
  \includegraphics[width=3.5in]{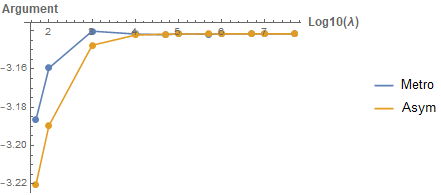}
  \end{minipage}%
  }%
  \centering
\caption{Absolute values and arguments of $\langle E^2_1\cdot E^3_1\rangle$. The results of the asymptotics are drawn in yellow line, and the numerical results are indicated by blue line.}\label{fig:EEA}
\end{figure}

\begin{figure}[htbp]
  \centering
  \subfigure[Absolute value versus $\lambda$]{
  \begin{minipage}[t]{0.5\linewidth}
  \centering
  \includegraphics[width=3.5in]{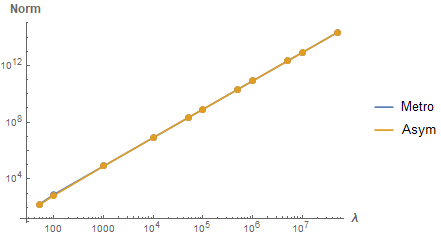}
  \end{minipage}%
  }%
  \subfigure[Argument versus $\lambda$]{
  \begin{minipage}[t]{0.5\linewidth}
  \centering
  \includegraphics[width=3.5in]{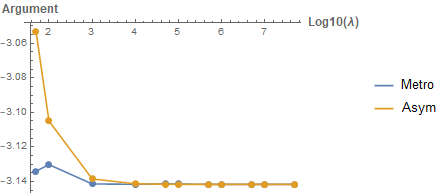}
  \end{minipage}%
  }%
  \centering
\caption{ Absolute values and arguments of $\langle E^1_4 \cdot E^5_4\rangle$. The results of the asymptotics are drawn in yellow lines, and the numerical results are indicated by the blue lines.}\label{fig:EEB}
\end{figure}

The expectation value of $\left\langle E^a_n\cdot E^b_n E^c_m\cdot E^d_m \right\rangle$ is a tensor with $1275$ non-zero components, and the expectation value of $\left\langle E^a_n\cdot E^b_n\right\rangle$ consists $50$ non-zero  components. We numerically compute these components depending on a sufficiently large number (over $10^7$) of samples obtained by the MCMC method. As an example of our computation, Figs. \ref{fig:EE}, \ref{fig:EEA}, and \ref{fig:EEB} plot the absolute values and the arguments of the components $\langle E^2_1\cdot E^3_1 E^1_4 \cdot E^5_4\rangle$, $\langle  E^1_4 \cdot E^5_4\rangle$ and $\langle E^2_1\cdot E^3_1\rangle$ respectively. In all these plots, the numerical results are shown on the blue lines, and the results given by the asymptotic expansion \eqref{eq:DEE1},  \eqref{eq:DEE2}, \eqref{eq:DEE3}, \eqref{eq:DEE4} are shown on the yellow lines. TABLE \ref{tab:EEdiff}, TABLE \ref{tab:E1diff} and TABLE \ref{tab:E2diff} record the percentage difference between the numerical results and the asymptotic results.

On the one hand, our results match the results from the asymptotics very well in the large spin limit. For the components we show here, the percentage differences between the asymptotic results and the numerical results are smaller than $0.03\%$ when $\lambda$ is greater than $10^5$, and the percentage differences tend to become smaller at larger $\lambda$. On the other hand, when $\lambda$ is small, the percentage differences become large because the higher order $1/\lambda$ corrections become important in this realm.

\begin{table}[htbp]
  \centering\caption{The difference between the numerically computed and asymptotically expanded $\langle E^2_1\cdot E^3_1 E^1_4 \cdot E^5_4\rangle$ }\label{tab:EEdiff}
	\small
	\setlength{\tabcolsep}{0.8mm}
	\begin{tabular}{|c|c|c|c|c|c|c|c|c|c|c|}
    \hline
		$\lambda$&$10^2$&$10^3$&$10^4$&$5\times10^4$&$10^5$&$5\times10^5$&$10^6$&$5\times 10^6$&$10^7$&$5\times 10^7$\\
		\hline
		Difference $(\%)$&8.71&0.79&0.12&0.052&0.036&0.017&0.0062&0.0018&0.00037&0.00069\\
		\hline
	\end{tabular}
\end{table}

\begin{table}[htbp]
  \centering\caption{The difference between the numerically computed and asymptotically expanded $\langle E^2_1\cdot E^3_1 \rangle$ }\label{tab:E1diff}
	\small
	\setlength{\tabcolsep}{0.8mm}
	\begin{tabular}{|c|c|c|c|c|c|c|c|c|c|c|}
    \hline
		$\lambda$&$10^2$&$10^3$&$10^4$&$5\times10^4$&$10^5$&$5\times10^5$&$10^6$&$5\times 10^6$&$10^7$&$5\times 10^7$\\
		\hline
		Difference $(\%)$&22.32&2.00&0.31&0.078&0.022&0.016&0.012&0.0022&0.000047&0.0016\\
		\hline
	\end{tabular}
\end{table}

\begin{table}[htbp]
  \centering\caption{The difference between the numerically computed and asymptotically expanded $\langle E^1_4 \cdot E^5_4 \rangle$ }\label{tab:E2diff}
	\small
	\setlength{\tabcolsep}{0.8mm}
	\begin{tabular}{|c|c|c|c|c|c|c|c|c|c|c|}
    \hline
		$\lambda$&$10^2$&$10^3$&$10^4$&$5\times10^4$&$10^5$&$5\times10^5$&$10^6$&$5\times 10^6$&$10^7$&$5\times 10^7$\\
		\hline
		Difference $(\%)$&18.66&1.18&0.18&0.026&0.017&0.00054&0.0037&0.00035&0.00036&0.00083\\
		\hline
	\end{tabular}
\end{table}

\subsection{Spinfoam Propagator}

\begin{figure}[h]
  \centering
  \subfigure[Absolute value versus $\lambda$]{
  \begin{minipage}[t]{0.5\linewidth}
  \centering
  \includegraphics[width=3.5in]{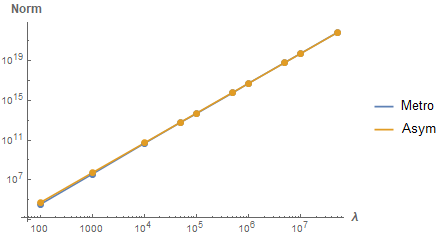}
  \end{minipage}%
  }%
  \subfigure[Argument versus $\lambda$]{
  \begin{minipage}[t]{0.5\linewidth}
  \centering
  \includegraphics[width=3.5in]{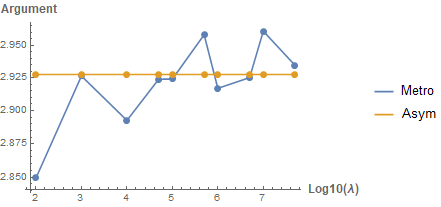}
  \end{minipage}%
  }%
  \centering
\caption{ Absolute values and arguments of $G^{2315}_{14}$ corresponding to different $\lambda$. The results of the asymptotics are drawn in yellow line, and the numerical results are indicated by blue line.}\label{fig:PCA}
\end{figure}

The propagator $G^{abcd}_{mn}$ also has $1275$ non-zero components. Figure \ref{fig:PCA} plots the absolute values and the arguments of the component $G^{2315}_{14}$. The percentage differences between the asymptotic limit (\ref{eq:DEE4}) and numerical results from the Lefschetz thimble Monte-Carlo are shown in TABLE \ref{tab:PCD}. For the results with $\lambda>10^6$, the percentage difference is smaller than $4\%$. This comparison shows that the asymptotic expansion and numerical Lefschetz thimble Monte-Carlo are consistent in the large spin limit. Similar to the computation of the expectation values, their differences become large in small spin realm because of the non-negligible contributions of the higher order $1/\lambda$ corrections.  

\begin{table}[h]
  \centering\caption{The difference between the numerically computed and asymptotically expanded $G^{2315}_{14}$ }\label{tab:PCD}
	\small
	\setlength{\tabcolsep}{0.8mm}
	\begin{tabular}{|c|c|c|c|c|c|c|c|c|c|c|}
    \hline
		$\lambda$&$10^2$&$10^3$&$10^4$&$5\times10^4$&$10^5$&$5\times10^5$&$10^6$&$5\times 10^6$&$10^7$&$5\times 10^7$\\
		\hline
		Difference $(\%)$&37.90&27.00&13.22&2.76&10.09&8.86&1.89&1.13&3.90&2.06\\
		\hline
	\end{tabular}
\end{table}

\begin{figure}[h]
  \centering
  \subfigure[$\lambda=10^6$ with different number of samples]{
  \begin{minipage}[t]{0.5\linewidth}
  \centering
  \includegraphics[width=3.5in]{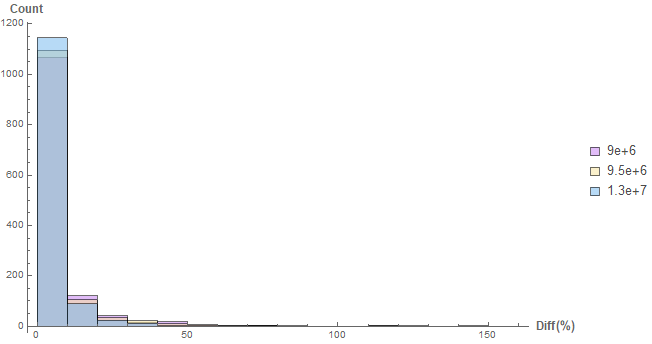}
  \end{minipage}%
  }%
  \subfigure[$\lambda=10^6$ v.s. $\lambda=10^7$]{
  \begin{minipage}[t]{0.5\linewidth}
  \centering
  \includegraphics[width=3.5in]{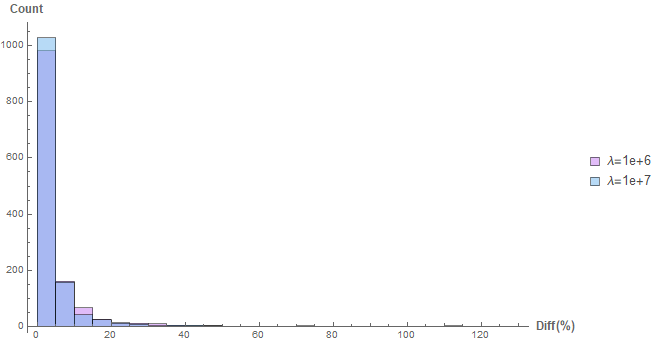}
  \end{minipage}%
  }%
  \centering
\caption{Histogram of the percentage errors of the components of $G^{abccd}_{mn}$ for (a)  $\lambda=10^6$ and (b) $\lambda=10^7$.}\label{fig:PCB}
\end{figure}

We compute all the $1275$ components of the propagator $G^{abcd}_{mn}$ and compare them with the results from the leading order asymptotics (\ref{eq:lead}). Figure \ref{fig:PCB}(a) shows the histograms of the percentage differences of the components of $G^{abcd}_{mn}$ between the asymptotic limit (\ref{eq:lead}) and the results from the Lefschetz thimble Monte-Carlo, at $\lambda=10^6$, based on $8985600$ samples, $9504000$ samples and $12787200$ samples. As we can see, the percentage differences for most of the components ($1067$ components for the result of $8985600$ samples, $1095$ components for the result of $9504000$ samples and $1144$ components for the result of $12787200$ samples) are smaller than $10\%$. There are however several components with percentage differences greater than $100\%$, one of which is nearly as much as $120\%$ in the  $12787200$ samples case. But a strong tendency that the percentage differences become smaller with respect to the increasing number of the samples (the maximum difference decreasing from $147\%$ in the $8985600$ samples cases to $120\%$ in the $16761600$ samples case) implicates that when $\lambda=10^6$, $12787200$ samples may not be enough to make the Markov chains perfectly converging to the desired distribution and cause such big differences. The percentage differences of these components will further decrease when the number of the samples increases.  

Figure \ref{fig:PCB} (b) draws a comparison between the histogram of the percentage differences of the components when $\lambda=10^7$ and when $\lambda=10^6$. The results are all achieved with $12787200$ samples. In the case of $\lambda=10^7$, the percentage differences of most of the components are less than $10\%$ and the maximum difference is around $45\%$. The comparison shows that the Markov chains converge to the desired distribution faster than they do in the case of $\lambda=10^6$, and the Lefschetz thimble Monte-Carlo results for $\lambda=10^7$ are more consistent to the asymptotic limit (\ref{eq:lead}). This fact might suggest that when $\lambda=10^7$, the less important $1/\lambda$ correction and the easier converging Markov chains are correlated.

In summary, the expectation values of the metric operators and the propagator obtained from Lefschetz thimble Monte-Carlo show their compatibility to the asymptotics from the stationary phase analysis in the large-$\lambda$ limit. As $\lambda$ increases, the compatibility to the asymptotics tends to be improved. These results fulfill our expectation about the semi-classical behavior of the spinfoam propagator and validates our algorithm and coding.


\section{Benchmarks}
\begin{table}[h]
  \centering\caption{The computer platforms}\label{tab:BCH}
	\small
	\setlength{\tabcolsep}{0.8mm}
	\begin{tabular}{|c|c|c|c|c|}
    \hline
    Platform&CPU&RAM&OS&Mathematica\texttrademark Version\\
    \hline
		(1)&AMD EPYC\texttrademark 7742x2 & 512G DDR4-3200 & Ubuntu\texttrademark 20.04.1 LTS & 12\\
		\hline
    (2)&AMD EPYC\texttrademark 7642x2 & 512G DDR4-3200 & Ubuntu\texttrademark 20.04.1 LTS &12\\
    \hline
    (3)&AMD Ryzen\texttrademark 3800XT & 32G DDR4-3200 & Windows\texttrademark 10 version 2004 & 12\\
		\hline
	\end{tabular}
\end{table}
We have tested our code \cite{hzcgit} in three different platforms shown in TABLE \ref{tab:BCH}. On platforms (1) and (2), the code runs with $54$ parallel Mathematica\texttrademark kernels, and for platform (3), the code runs with $16$ parallel Mathematica\texttrademark kernels. Both platforms (1) and (2) can update $60000$ samples per hour, and  platform (3) can update $20000$ samples per hour. 

\section{Conclusion}

In this paper, we propose a numerical method (combining the methods of Lefschetz thimble and DREAM) that can compute the expectation value of any observable in a quantum system described by a complex-valued action. We apply our method to compute the EPRL spinfoam propagator on a 4-simplex. Our computation focuses on the spinfoam propagator with relatively large spins $\lambda\geq 100$. Our results not only comply with the expected spinfoam semiclassical behavior but also gives rise to a quantum correction. The theory of Lefschetz thimble indicates that the quantum correction due to our computation sums $1/\lambda$ corrections to all orders. 

In principle, the method is applicable to all types of the spin foam model with Lorentzian or Euclidean signature, with different choices of the $Y$-map and with the different values of the Barbero-Immirzi parameter $\gamma$. Although, the explicit propagator components are different with respect to these choices, the conclusion that the numerical results comply with the semiclassical behavior holds for all these choices.

Our method is efficient and scalable for the numerical computation of the spinfoam model. We are capable of numerically computing spinfoams with relatively large spins. Our method can also compute the spinfoam model with multiple 4-simplices. The Lefschetz thimble and Monte-Carlo methods are able to compute oscillatory integrals of a few hundreds variables (see e.g. \cite{Alexandru:2018ngw,Alexandru:2020wrj} and the references therein). For instance, the EPRL spinfoam model (with the coherent intertwiner boundary) on the complex $\Delta_3$ of three 4-simplices and an internal face is a 133-(real)dimensional integral (see \cite{hanPI} for the integral formula on multiple 4-simplices), and thus can be handled by our method. Therefore, a future task is to compute the correlation functions of the spinfoam model on $\Delta_3$. An interesting aspect of the spinfoam model on $\Delta_3$ is its relation to the flatness problem: The spinfoam integral is dominated by the flat geometry (with vanishing deficit angle) when spins are large \cite{dona2020numerical,flatness,LowE1,frankflat,Perini:2012nd}. Hence, it is interesting to compute the spinfoam expectation value of the deficit angle on $\Delta_3$ and demonstrate its dependence on spins. Our method is suitable for computing expectation values and especially for studying their behaviors with large spins.

As another interesting future work, we can apply the Pachner 1-5 move and subdivide a 4-simplex into five 4-simplices, the spinfoam model on the resulting complex $\Delta_5$ is a 230-(real)dimensional integral, and is expected to be handled by our method. Our method can compute spinfoam expectation values and correlation functions on $\Delta_5$, compare them with the results on a 4-simplex, and understand their behaviors under the 1-5 move. Similar studies should be applied to the other elementary Pachner moves. The resulting behaviors should be useful for studying the renormalization of spinfoam model under changing triangulation. 

As an efficient way to compute oscillatory integrals, our method has wide applications and is not restricted to LQG/Spinfoam. In addition to Lattice Field Theory, where similar methods have been extensively applied, our method also applies to Topological Field Theory and Knot Theory. The analytically continued Chern-Simons theory and colored Jones polynomial are related to the finite dimensional integrals on Lefschetz thimbles. These finite dimensional integrals are known as holomorphic blocks \cite{Witten:2010cx,Beem:2012mb} (see also \cite{hanSUSY} for the relation with spinfoams). It is natural to apply our method to compute holomorphic blocks because they are defined on Lefschetz thimbles.




\section*{Acknowledgements}

Z.H. is supported by Xi De post-doc funding from State Key Laboratory of Surface Physics at Fudan University. M.H. receives support from the National Science Foundation through grant PHY-1912278. Y.W. is supported by NSFC grant No. 11875109 and Shanghai Municipal Science and Technology Major Project (Grant No.2019SHZDZX01). The authors thank Prof. Ling-Yan Hung  and Prof. Hui Luo for sharing their high performance computing resources.

\appendix

\section{Markov Chain Monte Carlo Methods}\label{apd:MCMC}
MCMC methods comprise a class of algorithms designed for sampling from a posterior probability distribution $\pi(x)$ on a given space $V$. In this appendix we review the main idea of the MCMC methods and several specific algorithms 
\subsection{Markov Chain Monte Carlo methods in general}
A Markov chain is defined as a stochastic model describing a sequence of random variates in which the probability distribution of each variate only depends on the the value of the previous one variate attained. 
For a Markov chain sequentially comprises $N$ sampled data, the Markov chain central limit theorem \cite{doi:10.1080/17442500500190060}\cite{brooks_meng_jones_galman_2011} guarantees that, when $N \to \infty$, the Markov chain will reach its equilibrium state so that sampled points by Markov chain will converge to a posterior probability distribution. 
An equilibrium chain must follow the Bayes local balance condition:
\be
\pi(y)K(x|y)=\pi(x)K(y|x),
\ee
where $\pi(x)$ is the posterior probability distribution and the transition kernel $K(x|y)$ is the conditional probability distribution of a random variate on the chain if its previous random variate is sampled as a value $y$. 

Imagine that a Markov chain as a `walker' moving on a phase space. Once a point being reached by the `walker', it is considered being sampled once. Then the Bayes local balance condition means that the marginal probability distribution of the `walker' first appears on $x$ then moves to $y$ is the same as the one that the `walker' appears on $y$ then moves to $x$. Thus the future movement of the 'walker' has no bias. The distribution of the sampled point will converge to the static probability distribution $\pi(x)$.

The idea of the MCMC integral method is to simulate such a `walker' randomly walking on phase space. If each step of the `walker's' random movement is designed to satisfy the transition kernel $K(x|y)$, after a long march, the history of `walker's' random movement will follow a posterior distribution, e.g., $\exp(S_{eff})$. Then, one can compute the mean value of the function $f(x)$ among the sampled history points to approximate $\langle f \rangle_{eff}$. 

\subsection{Metropolis algorithm}
Following the idea of MCMC, one important question is how to construct the transition kernel $K(x|y)$ so that the Markov chain can sample from a  
desired posterior distribution like $\exp(S_{eff})$. In Metropolis algorithm\cite{doi:10.1063/1.1699114} \cite[Chapter1.12]{brooks_meng_jones_galman_2011}, which is a type MCMC method, the transition kernel $K(x|y)$ is constructed as
\be\label{eq:metro}
K(x|y)=\alpha(x,y)p(x|y),
\ee
where $p(x|y)$ can be any proposal transition distribution and the acceptance rate $\alpha(x,y)$ is defined as
\be
\alpha(x,y)=\min \left\{1, \frac{\pi(x)p(y|x)}{\pi(y)p(x|y)} \right\}.
\ee
One can easily check the validation of the algorithm by plugging \eqref{eq:metro} back to the Bayes local balance condition. 

\subsection{Metropolis-Hastings algorithm}
As an improvement of the Metropolis algorithm, the Metropolis-Hastings (MH) algorithm uses symmetric proposal transition distributions to construct transition kernels $K(x|y)$. When a transition distribution is symmetric, $p(x|y)=p(y|x)$ and the acceptance rate $\alpha(x,y)$ reduces to
\be
\alpha(x,y)=\min \left\{1, \frac{\pi(x)}{\pi(y)} \right\}.
\ee

In details, the MH algorithm generate the samples in the steps shown in \textbf{Algorithm} \ref{alg:MH}
\begin{figure}[htbp]
\begin{algorithm}[H]
  \caption{Metropolis-Hastings (MH) algorithm}\label{alg:MH}
  \begin{algorithmic}[1]
    \STATE initial $x^{(0)}$
    \FOR{iteration $i=1,2,\cdots N$}
          \STATE Propose candidate $x^{cand}$ from $p(x|x^{(i-1)})$
          \STATE Acceptance rate $\alpha \gets \min \left\{1, \frac{\pi(x^{cand})}{\pi(x^{(i-1)})} \right\}$
          \STATE $u\sim$Uniform$(u;0,1)$
          \IF {$u<\alpha$}
              \STATE $x^{(i)} \gets x^{cand}$
          \ELSE
              \STATE $x^{(i)} \gets x^{(i-1)}$
          \ENDIF
    \ENDFOR
    \end{algorithmic}
\end{algorithm}
\end{figure}

When $N$ goes to be large, the sampled data $\{ x^{(i)} \}$ follows the posterior distribution $\pi(x)$. 

\subsection{Adaptive Metropolis-Hastings Algorithm}
The simplicity of the MH algorithm make it easy to be applied; however, the performance of the algorithm depends on the tuning of some internal variables, such as, the scale and orientation of the proposal distribution. On one hand, if the proposal distribution is too wide, most of the candidates will be rejected and the chain's convergence to the target distribution will be delayed. On the other hand, if the proposal distribution is too narrow, although most of the candidate will be accept, it may take a very large number of updates to make the Markov chain move to the most probable region and converge to the target distribution. One improved MH is the Adaptive Metropolis Hastings (AM) algorithm \cite{Haario1999,haario2001}. The AM algorithm uses single Markov chain, but it is a powerful algorithm that can automatically select the appropriate proposal distribution. For a $d$-dimensional phase space, by using the sampled data, AM continuously adapt the covariance, denoted by $C_t$, of the Gaussian proposal distribution $p(x|x^{(i-1)})=N(x^{(i_1)},C_t)$. Explicitly, 
\be
C_t=S_d Cov(x^{(0)},\cdots,x^{(i-1)})+\epsilon I_d,
\ee
where the $S_d$ is the scaling factor depending on $d$, $I_d$ is the $d$-dimensional identity matrix, and $\epsilon$ is a very small data comparing to the scale of $\{ x^{(i)} \}$. If the target distribution is a Gaussian distribution, the optimal $S_d$ is $2.4^2/d$ so that the acceptance rate $\alpha$ can stay around $0.3$. The detailed steps of AM (\textbf{Algorithm} \ref{alg:AM}) is similar to MH.

\begin{figure}[htbp]
  \begin{algorithm}[H]
    \caption{Adaptive Metropolis-Hastings (AM) algorithm}\label{alg:AM}
    \begin{algorithmic}[1]
      \STATE initial $x^{(0)}$, $C_t \gets I_d$
      \FOR{iteration $i=1,2,\cdots N$}
            \STATE Propose candidate $x^{cand}$ from $N(x^{(i_1)},C_t)$
            \STATE Acceptance rate $\alpha \gets \min \left\{1, \frac{\pi(x^{cand})}{\pi(x^{(i-1)})} \right\}$
            \STATE $u\sim$Uniform$(u;0,1)$
            \IF {$u<\alpha$}
                \STATE $x^{(i)} \gets x^{cand}$
            \ELSE
                \STATE $x^{(i)} \gets x^{(i-1)}$
            \ENDIF
            \STATE $C_t \gets S_d Cov(x^{(0)},\cdots,x^{(i)})+\epsilon I_d$
      \ENDFOR
      \end{algorithmic} 
  \end{algorithm}
  \end{figure}

\subsection{Differential evolution Markov Chain algorithm}
The AM algorithm works fine for many simple inference problems, but its efficiency becomes low when dealing with complicated posterior distribution, especially for high-dimensional problems. In order to deal with these limitations, Differential evolution Markov chain (DE-MC) \cite{Braak2006} is developed. In DE-MC, $M$ different Markov chains $\{ x_s^{(t)}, s=1,\cdots, M \}$ are run in parallel. In stead of using the covariance of the previoused samples, i.e.,$Cov(x^{(0)},\cdots,x^{(i-1)})$, DE-MC uses the current location of the chains to generate the candidates. At each updating step, instead of using proposal distribution to generate candidates, DE-MC uses a genetic algorithm, called Differential evolution algorithm, to generate the the candidates as
\be
x_s^{cand}=x_s^{(t-1)}+\gamma (x_{s_1}^{(t-1)}-x_{s_2}^{(t-1)})+\epsilon,\ s=1,\cdots M, 
\ee
where $\gamma$ is a scaling factor, $s_1$ and $s_2$ are labels of another two chains different from the chain $s$, and $\epsilon$ is a number draw from the uniform distribution Uniform$(-b,b)$ with $|b|<1$. Similar to the AM algorithm, when the posterior distribution is $d$-dimensional Gaussian, the optimal choice of $\gamma$ is $2.4/\sqrt{2d}$. For every $10$ update steps, $\gamma=1$ to allow a direct jumps between the chains. When dealing with multimodal posterior distribution, this brings in a huge of advantage comparing to single chain AM algorithm. In AM algorithm, it is hard to make the Markov chain tunnel between two distinct possible regions, while in DE-MC, different chains can `explore' different regions simultaneously and the jumps between the regions are allowed so that the samples from different regions are well mixed to satisfy the target posterior distribution. The steps of the DE-MC algorithm is shown in \textbf{Algorithm} \ref{alg:DE-MC}.

\begin{figure}[htbp!]
\begin{algorithm}[H]
  \caption{DE-MC algorithm}\label{alg:DE-MC}
  \begin{algorithmic}[1]
    \STATE initial $x_s^{(0)}$, $s=1, \cdots, M$
    \FOR{iteration $i=1,2,\cdots N$}
      \FOR{chains $j=1, \cdots, M$}
        \WHILE{$s_1=j$}
          \STATE $s_1\gets$ RandomInteger$(1,M)$
        \ENDWHILE
        \WHILE{$s_2=j$ or $s_2=s_1$}
          \STATE $s_2\gets$ RandomInteger$(1,M)$
        \ENDWHILE
        \STATE Propose candidate $x^{cand}\gets x_{j}^{(i-1)}+\gamma (x_{s_1}^{(i-1)}-x_{s_2}^{(i-1)})+\epsilon$
        \STATE Acceptance rate $\alpha \gets \min \left\{1, \frac{\pi(x^{cand})}{\pi(x_j^{(i-1)})} \right\}$
        \STATE $u\sim$Uniform$(u;0,1)$
            \IF {$u<\alpha$}
                \STATE $x_j^{(i)} \gets x_j^{cand}$
            \ELSE
                \STATE $x_j^{(i)} \gets x_j^{(i-1)}$
            \ENDIF
      \ENDFOR    
    \ENDFOR
  \end{algorithmic}
\end{algorithm}
\end{figure}
The DE-MC is efficiently accommodate to the situation when the target posterior distribution is complicated, and high-dimensional. The efficiency can be further enhanced by making several modifications on the algorithm. These modifications brings in the Differential Evolution Adaptive Metropolis (DREAM) algorithm used in our paper to compute the propagator.

\newpage
\bibliographystyle{apsrev}
\bibliography{graviton}
\end{document}